\documentclass[lettersize,journal]{IEEEtran}
\IEEEoverridecommandlockouts
\usepackage{cite}
\usepackage{bm}
\usepackage{amsmath,amssymb,amsfonts}
\usepackage{algorithm}
\usepackage{algpseudocode}
\usepackage{graphicx}
\usepackage{amsfonts}
\usepackage{amssymb}
\usepackage{textcomp}
\usepackage[T1]{fontenc}
\usepackage{longtable}
\usepackage{supertabular}
\usepackage{xcolor}
\usepackage{subcaption}
\usepackage{tabularx}
\usepackage{comment}
\usepackage{booktabs}
\usepackage{mathrsfs} 
\usepackage{float}
\usepackage{booktabs}
\usepackage{multirow}
\newtheorem{definition}{Definition}
\newtheorem{theorem}{Theorem}

\newtheorem{remark}[theorem]{Remark}

\allowdisplaybreaks
\def\BibTeX{{\rm B\kern-.05em{\sc i\kern-.025em b}\kern-.08em
    T\kern-.1667em\lower.7ex\hbox{E}\kern-.125emX}}

\allowdisplaybreaks
\def\BibTeX{{\rm B\kern-.05em{\sc i\kern-.025em b}\kern-.08em
    T\kern-.1667em\lower.7ex\hbox{E}\kern-.125emX}}
\begin{document}

\title{
Enhancing System-Level Safety in Mixed-Autonomy Platoon via Safe Reinforcement Learning

\author{Jingyuan Zhou, Longhao Yan, and Kaidi Yang%
\thanks{The authors are with the Department of Civil and Environmental Engineering, National University of Singapore, Singapore 119077. Email:{\{jingyuanzhou, longhao.yan\}@u.nus.edu, 
 kaidi.yang@nus.edu.sg}.}
 \thanks{This research was supported by the Singapore Ministry of Education (MOE) under NUS Start-Up Grant (A-8000404-01-00). This article
solely reflects the opinions and conclusions of its authors and not Singapore MOE or any other entity.}
 
 }}

\maketitle

\begin{abstract}
Connected and automated vehicles (CAVs) have recently gained prominence in traffic research due to advances in communication technology and autonomous driving. Various longitudinal control strategies for CAVs have been developed to enhance traffic efficiency, stability, and safety in mixed-autonomy scenarios. Deep reinforcement learning (DRL) is one promising strategy for mixed-autonomy platoon control, thanks to its capability of managing complex scenarios in real time after sufficient offline training. However, there are three research gaps for DRL-based mixed-autonomy platoon control: (i) the lack of theoretical collision-free guarantees, (ii) the widely adopted but impractical assumption of skilled and rational drivers who will not collide with preceding vehicles, and (iii) the strong assumption of a known human driver model. To address these research gaps, we propose a safe DRL-based controller that can provide a system-level safety guarantee for mixed-autonomy platoon control. First, we combine control barrier function (CBF)-based safety constraints and DRL via a quadratic programming (QP)-based differentiable neural network layer to provide theoretical safety guarantees. Second, we incorporate system-level safety constraints into our proposed method to account for the safety of both CAVs and the following HDVs to address the potential collisions due to irrational human driving behavior. Third, we devise a learning-based system identification approach to estimate the unknown human car-following behavior in the real system. Simulation results demonstrate that our proposed method effectively ensures CAV safety and improves HDV safety in mixed platoon environments while simultaneously enhancing traffic capacity and string stability.
\end{abstract}


\section{Introduction}
\IEEEPARstart{C}{onnected}  and automated vehicles (CAVs) have been widely recognized to be beneficial for traffic flow, thanks to their capability of exchanging real-time information and eliminating undesirable human driving behavior~\cite{deng2023cooperative,li2017dynamical,tilg2018evaluating,zhou2022safety,liang2023robust}. Significant research efforts have been placed on longitudinal control, i.e., the coordinated control of the speed profile of a platoon of CAVs, with enormous potential to enhance road capacity \cite{sala2021capacity}, traffic stability \cite{wang2021leading,chen2022robustly}, energy efficiency \cite{tsugawa2016review,li2021reinforcement,yang2022eco}, and safety \cite{xiao2021decentralized,zhou2022safety,zhao2023safety,liu2023reachability,liu2022structural,10422463}. Early research on longitudinal control tends to assume all vehicles to be CAVs~\cite{dey2015review,milanes2013cooperative}. 
However, such an assumption is not practical in the near future since the penetration rates of CAVs can only increase gradually as the technology matures and public acceptance improves. Therefore, it is crucial to investigate mixed-autonomy platoons, whereby CAVs and human-driven vehicles (HDVs) coexist.

Several control schemes have been proposed in the context of mixed-autonomy platoons, whereby CAVs are controlled considering the behavior of HDVs around them. One typical scheme is connected cruise control~\cite{orosz2016connected,shen2023energy}, whereby the CAV at the tail of a platoon adjusts its driving decisions considering multiple HDVs ahead. However, such a scheme only considers the preceding vehicles of the studied CAV without accounting for the CAV's impact on the following vehicles. In contrast, the recently proposed scheme of leading cruise control (LCC) extends traditional cruise control by allowing CAVs to utilize information from both preceding and following HDVs~\cite{wang2021leading}. With the additional information and flexibility compared to connected cruise control, LCC offers a better potential to improve the performance of the entire platoon. 

Various controllers have been developed to implement LCC, including linear feedback controllers \cite{wang2021leading}, optimization-based controllers using model predictive control (MPC) \cite{feng2021robust,zhan2022data} and data-enabled predictive control (DeePC)~\cite{chen2021mixed,wang2022data,zhou2024parameter}, and learning-based controllers based on deep reinforcement learning (DRL)~\cite{shi2021connected,shi2023deep}. Linear feedback controllers are easy to design but can hardly handle complex constraints and optimization objectives. MPC can explicitly incorporate complex constraints and objectives into an embedded optimization problem, which, however, relies on explicit system modeling and requires accurate system identification to estimate system parameters. Although DeePC can simultaneously perform system identification and control in a data-driven nonparametric manner, it may require a large amount of computational resources to update the decision variables in each decision step. Among all these controllers, DRL-based controllers appear promising in that they can handle complex scenarios with a marginal online computational burden after sufficient offline training.

However, existing research on DRL-based mixed-autonomy platoon control suffers from three limitations. First, safety is only indirectly considered by incorporating safety-related penalties into the reward function, which can hardly provide safety guarantees due to the black-box nature of RL. 
Second, existing works tend to assume skilled and rational human drivers who will not collide with their preceding vehicles, and consequently only consider ego-vehicle safety \cite{wang2023deep, wang2022distributed}, i.e., the safety spacing between the controlled CAV and its preceding vehicle. However, such an assumption is not realistic since human errors are common. The undesirable behavior of the following HDVs may place other vehicles in the platoon in dangerous situations and thereby undermine \emph{system-level} safety for both HDVs and CAVs. Moreover, we recognize that CAVs, once detecting such undesirable behavior of HDVs, have the potential to adjust their acceleration to improve the safety performance of the entire platoon. Therefore, it is beneficial to consider system-level safety when controlling CAVs. Third, existing literature often assumes the human driver model to be known to CAVs, which, nevertheless, does not hold in real traffic systems, where driver behavior can be time-varying and diverse.

To address these limitations, we propose a safe DRL-based controller that can provide a system-level safety guarantee for mixed-autonomy platoon control in a single-lane freeway section. Our work extends \cite{xiao2023barriernet} that incorporated safety guarantees into a supervised learning-based controller via a quadratic programming (QP)-based safety filter, whereby the safety constraints are characterized by control barrier functions (CBF)~\cite{ames2014control}. It should be noted that the canonical form of \cite{xiao2023barriernet} is not capable of addressing the aforementioned limitations, and hence we extend \cite{xiao2023barriernet} from the following three perspectives. First, \cite{xiao2023barriernet} leveraged supervised learning to imitate an optimal controller whose decisions are pre-calculated and used as labels, which does not fit our case where labels are unavailable. We address this limitation by devising a differentiable safety module for the DRL-based controller in an unsupervised manner to provide safety guarantees in both online training and offline testing processes. Second, applying CBF to protect system-level safety requires the use of high-order CBFs \cite{xiao2021high}, which involves finding high-order derivatives and can be computationally expensive and imprecise in real systems. To address this issue, we introduce reduced order CBF candidates to make the relative degree for CBF candidates of the following vehicles remain $1$ to avoid calculating high-order derivatives. Third, note that constructing CBF requires the explicit modeling of HDVs' system dynamics, which is nevertheless unknown in the real traffic scenario. To address the issue, we employ a learning-based human driver behavior identification approach to characterize the unknown car-following behavior.

\emph{Statement of contribution}. The contributions of this paper are three-fold. First, we design a safe DRL-based control strategy for mixed-autonomy platoons by combining CBF and DRL via a QP-based differentiable neural network layer. This method not only enhances traffic efficiency by ensuring robust training performance but also provides safety guarantees throughout the entire training and testing processes of the DRL algorithm. Second, unlike existing works that consider only ego-vehicle safety with the underlying assumption of rational and skilled human drivers, we incorporate system-level safety constraints into DRL-based mixed-autonomy platoon control methods to account for the safety of both CAVs and the following HDVs. Third, to address the issue of lacking explicit HDV dynamics for constructing CBFs, we implement a learning-based system identification approach that allows us to estimate the unknown human car-following behavior in the real-world system.

A preliminary version of this paper \cite{10422463} was presented at the 2023 Intelligent Transportation Systems Conference. In this full version, we have made the following extensions: (i) generalizing the proposed methods from considering solely CAV safety to encompass system-level safety for both CAVs and HDVs to handle potentially undesirable HDV behavior, (ii) introducing an online learning-based system identification approach to estimate the unknown car-following dynamics of surrounding HDVs, and (iii) presenting new simulation results, e.g., safety regions for both CAVs and HDVs, to better illustrate the efficacy of our proposed method in enhancing safety performance.

The rest of this paper is organized as follows. Section~\ref{Preliminaries} introduces preliminaries about RL and CBFs. Section~\ref{Mixed Traffic Modelling and Controller Design} presents the system modeling for mixed-autonomy platoon control. Section~\ref{Safety-Critical Control of Mixed Platoon} presents our methodological framework for LCC that integrates safety guarantees into DRL via a CBF-QP-based approach and the online learning-based system identification formulation. Section~\ref{Simulation Results} conducts simulations to evaluate our proposed method. Section~\ref{Conclusion} concludes the paper.

\section{Preliminaries}
\label{Preliminaries}
In this section, we introduce Reinforcement Learning (RL) (Section~\ref{subsec:RL}) and Control Barrier Function (CBF) (Section~\ref{subsec:CBF}) as theoretical foundations for the proposed approach. 

\subsection{Reinforcement Learning (RL)} \label{subsec:RL}
RL addresses the problem of an intelligent agent learning to make decisions from its interactions with a dynamic environment. Specifically, the decision-making process is modeled as a Markov decision process (MDP) $\mathcal{M}=(\mathcal{X}, \mathcal{U}, P, r, \gamma)$, where $\mathcal{X}$ and $\mathcal{U}$ denote the set of states and the set of actions, respectively, and $P:\mathcal{X} \times \mathcal{U} \to \mathcal{X}$ represents the system dynamics in the form of transition probabilities $P(\bm{x}_{t+1}|\bm{x}_t,\bm{u}_t)$ with states $\bm{x}_t,\bm{x}_{t+1} \in \mathcal{X}$ and action $ \bm{u}_t\in \mathcal{U}$. 
The reward function $r: \mathcal{X}\times \mathcal{U} \to \mathbb{R}$ defines the reward collected from the environment by performing action $\bm{u}_t$ when the state is $\bm{x}_t$, and $\gamma\in (0,1]$ refers to the discount factor for future reward. The goal of RL is to learn a policy $\pi:\mathcal{X}\to \mathcal{U}$ in the form of a conditional probability $\pi(\bm{u}_t|\bm{x}_t)$ with $\bm{x}_t \in \mathcal{X}$ and $\bm{u}_t\in \mathcal{U}$ to maximize the discounted reward under this policy, which can be written as $\max~J(\pi) = \mathbb{E}_{\kappa\sim p_{\pi}}\left[\sum_{t=0}^T\gamma^{t}r(\bm{x}_t,\bm{u}_t)\right]$, where $\kappa=(\bm{x}_0,\bm{u}_0,\cdots,\bm{x}_T,\bm{u}_T)$ represents a trajectory defined as the sequence of states and actions of length $T$, and $p_{\pi}$ denotes the distribution of trajectories under policy $\pi$. 

In this paper, Proximal Policy Optimization 
 (PPO) is utilized as the fundamental RL algorithm to train a policy for CAVs. It is constructed in an actor-critic structure, whereby an actor network $\pi_{\theta_{\text{RL}}}$ produces actions based on the observed state, and a critic network $V_\phi(\bm{x}_t)$ estimates the state-action value of the MDP and generates loss values during agent training.

During the training process, the critic network updates its parameters with the aim of minimizing the error between the predicted value function and the actual return:
\begin{equation}
\mathcal{L}(\phi)=\mathbb{E}_{\bm{x}_t, \bm{u}_t \sim \pi_{\theta_{\text{RL}}}}\left[\delta^2\right], \quad \delta=r+\gamma V_\phi\left(\bm{x}_{t+1}\right)-V_\phi(\bm{x}_t)
\end{equation}
where $\mathcal{L}(\phi)$ represents the loss function of critic network, and $\delta$ is the temporal-difference error to be approximated. 

The actor network then gets updated following the loss function evaluated by the critic network:
\begin{equation}
\begin{aligned}
&L^{\operatorname{clip}}(\pi_{\theta_{\text{RL}}})=\\
&\mathbb{E}_{\bm{x}_t, \bm{u}_t \sim \pi_{\theta_{\text{RL},\text {old }}}}\left[\operatorname { m i n } \left(\frac{\pi_{\theta_{\text{RL}}}(\bm{u}_t \mid \bm{x}_t)}{\pi_{\theta_{\text{RL},\text {old }}}(\bm{u}_t \mid \bm{x}_t)} A^{\pi_{\theta_{\text{RL},\text {old }}}}(
\bm{x}_t, \bm{u}_t), \right.\right.
\\&\left.\left.\operatorname{clip}\left(\frac{\pi_{\theta_{\text{RL}}}(\bm{u}_t \mid \bm{x}_t)}{\pi_{\theta_{\text{RL},\text {old }}}(\bm{u}_t \mid \bm{x}_t)}, 1-\epsilon, 1+\epsilon\right) A^{\pi_{\text{RL},\theta_{\text {old }}}}(\bm{x}_t, \bm{u}_t)\right)\right]
\end{aligned}
\end{equation}
The function $\text{clip}(\cdot)$ prevents aggressive updating by utilizing a trust region defined by a hyperparameter $\epsilon$. $ A^{\pi_{\theta_{\text{RL},\text {old }}}}(\bm{x}_t, \bm{u}_t)$ is the advantage function that evaluates the benefit of selected actions.
\subsection{Control Barrier Functions}\label{subsec:CBF}
We next introduce Control Barrier Functions that characterize safety conditions into the constraints of control input. Specifically, consider a control affine system:
\begin{equation}
    \dot x=f(x)+g(x)u,
    \label{affine}
\end{equation}
where $x \in D \subset \mathbb{R}^{n}$, $f: \mathbb{R}^{n} \rightarrow \mathbb{R}^{n}$ and $g: \mathbb{R}^{n} \rightarrow \mathbb{R}^{n\times m}$ are locally Lipschitz continuous, $u \in U \subset \mathbb{R}^{m}$. 

\begin{definition}[Control Barrier Function~\cite{ames2014control}] Let $\mathcal{C} \subset D \subset \mathbb{R}^{n}$, with safe set $\mathcal{C}=\left\{x \in D \subset \mathbb{R}^{n}: h(x) \geq 0\right\}$ be the superlevel set of a continuously differentiable function $h:D \rightarrow \mathbb{R}$, then $h$ is a control barrier function for system \eqref{affine} if there exists an extended class $\mathcal{K}_\infty$  functions\footnote{As in \cite{rugh1981nonlinear}, a continuous function $\alpha_{\text{CBF}}$ is said to belong to extended class $\mathcal{K}_{\infty}$ if it is strictly increasing and $\alpha_{\text{CBF}}(0) = 0$} $\alpha_{\text{CBF}}$ such that:
\label{cbf}
\begin{equation}
    \sup _{u \in U}\left[L_{f} h(x)+L_{g} h(x) u\right] \geq-\alpha_{\text{CBF}}(h(x)), \forall x\in D
    \label{CBF definition}
\end{equation}
where $L_{f}h(x)$ and $L_{g}h(x)$ are the Lie derivatives of CBF candidate $h(x)$ along system dynamics $f(x)$ and $g(x)$, i.e., $\displaystyle L_{f}h(x)=\Big(\frac{\partial h(x)}{\partial x}\Big)^Tf(x)$ and $\displaystyle L_{f}g(x)=\Big(\frac{\partial h(x)}{\partial x}\Big)^Tg(x)$, which describe the rate of change of a tensor field (i.e. $h(x)$) along the flow generated by a vector field (i.e. $f(x),~g(x)$). 
\end{definition}

Eq. \eqref{CBF definition} for CBF $h$ implies that there exists control input $u$ such that $\dot{h} = \frac{\partial h(x)}{\partial x}\dot{x} = L_{f} h(x)+L_{g} h(x) u \geq -\alpha_{\text{CBF}}(h)$. Hence, CBF $h$ defines a forward-invariant safe set $\mathcal{C}$ \cite{aubin2011viability}, which ensures that if the system's current state lies within the safe set $\mathcal{C}$ and the control input adheres to the constraints imposed by the CBF, then the subsequent states are guaranteed to remain within the safe set, thereby ensuring the system's safety over time. 

\section{Mixed-Autonomy Traffic Environment Modeling}
\label{Mixed Traffic Modelling and Controller Design}
We consider a mixed-autonomy platoon in one lane of a freeway segment (either straight or curved). This platoon comprises HDVs (denoted by set $\Omega_{\mathcal{H}}$) and CAVs (denoted by set $\Omega_{\mathcal{C}}$), whereby the driving behavior of HDVs is modeled using car-following models (unknown to CAVs) and CAVs are controlled by RL-based controllers. CAV $i\in\Omega_{\mathcal{C}}$ collects state information within the communication range from both preceding and following vehicles in sets $\Omega_{\mathcal{P},i}$  and $\Omega_{\mathcal{F},i}$, respectively, and determines its control action based on the collected information. Moreover, we envision that the CAV penetration rate will remain very low at the early stage of CAV deployment, and hence it would be common for a CAV to be surrounded by multiple HDVs. Therefore, we assume that only one CAV is present in the platoon, i.e., $|\Omega_{\mathcal{C}}|=1$ with $|\cdot|$ denotes the cardinality of a set.

For each vehicle $i\in \Omega_{\mathcal{C}} \cup \Omega_{\mathcal{H}}$, we consider the dynamics described by second-order ordinary differentiable equations with states including the speed of vehicle $i$ (i.e., $v_i$) and the spacing between vehicle $i$ and vehicle $i-1$ (i.e., $s_i$). 

Specifically, the dynamics of CAVs can be described as:
\begin{equation}
\left\{\begin{array}{l}
    \dot{s}_{i}(t) =v_{i-1}(t)-v_{i}(t),\\
    \dot{v}_{i}(t) =u_i(t),
    \end{array} \quad i \in \Omega_\mathcal{C},\right.
    \label{CAV_sv}
\end{equation}
where the acceleration rate is the control input $u_i(t)$.

The dynamics of HDVs can be described as:
\begin{equation}
\left\{\begin{array}{l}
    \dot{s}_i(t) = v_{i-1}(t)-v_{i}(t),\\
    \dot{v}_i(t)=\mathbb{F}\left(s_i(t), v_i(t), v_{i-1}(t)\right),
\end{array} \quad i \in \Omega_\mathcal{H},\right.
\label{HDV_sv}
\end{equation}
where the acceleration rate  is determined by a car-following model $\mathbb{F}(\cdot)$ as a function of the spacing $s_i(t)$, the velocity of the preceding vehicle $v_{i-1}(t)$, and its own velocity $v_i(t)$.  Note that unlike existing works \cite{wang2021leading} that assume the model $\mathbb{F}(\cdot)$ to be public knowledge, we make a realistic assumption that both the formulation and parameters of $\mathbb{F}(\cdot)$ remain unknown to the CAV, and the CAV can only learn $\mathbb{F}(\cdot)$ from its observations.

Combing the dynamics of CAVs~\eqref{CAV_sv} and HDVs~\eqref{HDV_sv}, the longitudinal dynamics of the mixed-autonomy platoon can be written as:
\begin{equation}
    \dot{x}(t)=f(x(t),v_0(t))+B u(t),
    \label{continuous system}
\end{equation}
where $x(t)=\left[ s_{1}(t), v_{1}(t), \ldots, s_{n}(t), v_{n}(t)\right]^{\top}$ denotes the system states of all CAVs and HDVs, $v_0(t)$ denotes the velocity of the head vehicle, and $f\left(\cdot\right)$ indicates vehicle dynamics including the nonlinear dynamics of HDVs and the linear dynamics of CAVs. System matrix $B$ for CAVs' control input is summarized as $
    B=\left[e_{2 n}^{2 i_1}, e_{2 n}^{2 i_2}, \ldots, e_{2 n}^{2 i_m}\right] \in \mathbb{R}^{2 n \times m}$, 
where $n$ and $m$ represent the number of vehicles (including CAVs and HDVs) and the number of CAVs, respectively, and the vector $e^i_{2n}\in\mathbb{R}^{2n}$ associated with CAV $i$ is a vector with the $2i$-th entry being 1 and the others being 0. 

\section{Safety-Critical Learning-based Control for Mixed-Autonomy Platoons}
\label{Safety-Critical Control of Mixed Platoon}
\begin{figure*}[h]
    \centering
    \includegraphics[width = 15cm]{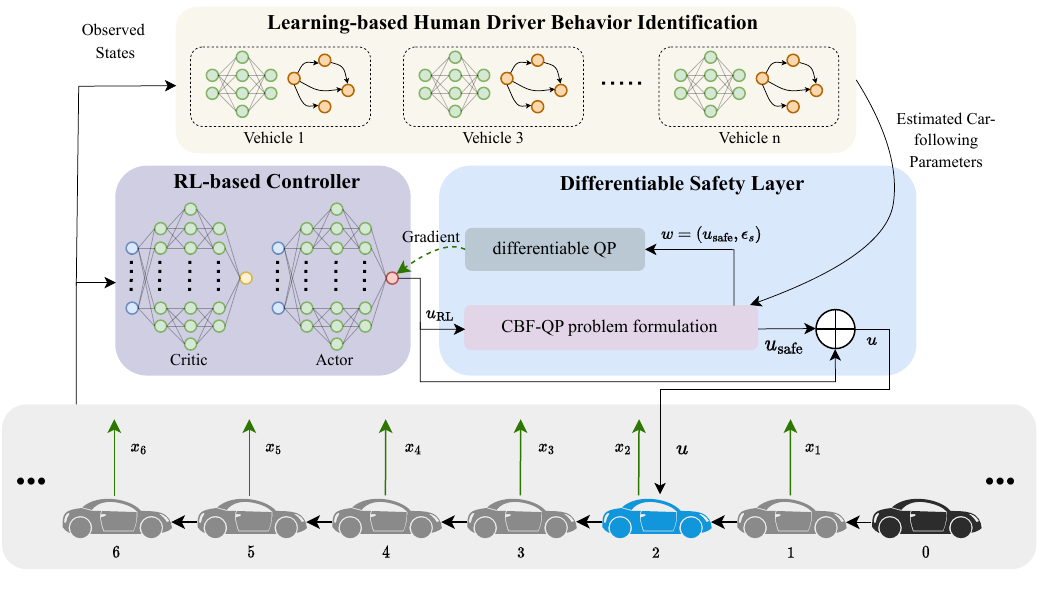}
    \caption{Overview of the proposed controller for CAVs in mixed-autonomy traffic framework, the dotted line represents backpropagation of the differentiable QP.}
    \label{fig:CAV_overview}
\end{figure*}
The overall framework of the proposed method is shown in Fig.~\ref{fig:CAV_overview}. The method involves three modules: an RL module, a learning-based system identification module, and a differentiable safety module. For each CAV (e.g., the blue car in Fig.~\ref{fig:CAV_overview}), the RL module takes its observed states as input and gives an output of the control decision $u_{\text{RL}}$. Note that although the RL reward function can include safety-related components, there is no guarantee that $u_{\text{RL}}$ is safe due to the black-box nature of RL. The differentiable safety module aims to provide a safety guarantee by converting $u_{\text{RL}}$ into a safe action. Specifically, a residual term $u_{\text{safe}}$ is calculated by solving a quadratic programming (QP) problem with CBF-based constraints (i.e., CBF-QP problem in short) to compensate  $u_{\text{RL}}$ such that the final action $u=u_{\text{RL}}+u_{\text{safe}}$ is a safe action. However, despite the CBF-QP problem being able to provide a safety guarantee, it cannot guide the training of RL, i.e., providing gradient information into the backpropagation process. To tackle this issue, the differentiable safety module further incorporates a differentiable QP layer to calculate and backpropagate the gradients of the obtained QP solution (i.e., $u_{\text{safe}}$) with respect to $u_{\text{RL}}$ by performing local sensitivity analysis on the KKT conditions of the QP. With such gradients, the DRL can improve its performance under safety constraints and expedite exploration. Moreover, constructing CBF constraints requires the knowledge of surrounding HDVs' car-following behavior, which is unfortunately unknown in real traffic scenarios. To address this issue, we develop a learning-based human driver behavior identification module that takes observed states as input to estimate the car-following models of surrounding HDVs. 

We next present the details of the three modules. Section~\ref{subsec:RL controller} presents the RL-based controller, Section~\ref{subsec:SI} provides the details of the learning-based human driver behavior identification module, and Section~\ref{subsec:QP} presents the CBF-QP module to incorporate safety guarantees into DRL via a differentiable QP layer.

\subsection{RL-based Controller} \label{subsec:RL controller}

We formulate the control of mixed-autonomy platoons as an MDP $\mathcal{M} = (\mathcal{X}, \mathcal{U}, P, r, \gamma)$, where $\mathcal{X}$ includes all possible states $x(t)$ defined as the spacing and velocity, $\mathcal{U}$ comprises all possible actions $u(t)$ defined as the acceleration rate of the CAV, the transition function $P$ (i.e., system dynamics) is defined in Eq.~\eqref{continuous system} that involves the system function $f(x(t),v_0(t))$ and control matrix $B$, and the policy of the RL-based controller is parameterized by $\theta_{\text{RL}}$. The reward function $r$ is introduced next. 

Our control objective of each CAV in mixed-autonomy platoons is to stabilize traffic flow, improve traffic efficiency, and ensure its own safety. Therefore, the reward function of CAV $i$ contains three parts:
\begin{equation}
    R_i = w_{\text{stability}}r_{\text{stability},i} + w_{\text{efficiency}}r_{\text{efficiency},i} + w_{\text{safety}}r_{\text{safety},i}
\end{equation}
where $R_i$ represents the total reward, $r_{\text{stability},i}$, $r_{\text{efficiency},i}$, and $r_{\text{safety},i}$  denote the string stability reward, efficiency reward, and safety reward for the CAV at position $i$, respectively. The weighting coefficients $w_{\text{stability}}$, $w_{\text{efficiency}}$, and $w_{\text{safety}}$ are designed by the CAV manufacturer or operator to reflect their preferences over these control objectives. 
Next, these three types of rewards will be discussed in detail.

For stability, we focus on string stability~\cite{feng2019string,liu2022structural}, a desirable property of vehicle platoons such that disturbances of the head vehicle do not get amplified when propagating through the platoon.
Specifically, we aim to minimize velocity oscillations of both the CAV and the following vehicles resulting from the disturbances of the preceding vehicle $i-1$: 
\begin{equation}
    r_{\text{stability},i} = -(v_i-v_{i-1})^2 - \sum_{j \in \Omega_\mathcal{F}} \kappa_j(v_j - v_{i-1})^2
\end{equation}
where $\kappa_j\leq 1$ is the coefficient for the following vehicle velocity oscillation. In the current implementation, we have set each weighting coefficient to 1, following \cite{wang2023deep}.

We characterize traffic efficiency as follows~\cite{vogel2003comparison}:
\begin{align}
    r_{\text{efficiency},i}=\left\{\begin{array}{lc}
                                -1, & \mathrm{TH}_i \geq 2.5 \\
                                0, & \text { otherwise }
                                \end{array}\right.
\end{align}
where $TH_i(t)=\frac{s_i(t)}{v_i(t)}$ approximates the headway. The goal is to increase flow represented as the reciprocal of headway. The headway threshold is set to ensure high flow on the road section.

For safety, we follow the common practice in existing works (e.g., \cite{wang2022adaptive}) and design 
a reward related to the time-to-collision (TTC)  metric. 
\begin{align}
&r_{\text{safety},i}= \begin{cases}\log \left(\frac{\mathrm{TTC}_i}{4}\right), & 0 \leq \mathrm{TTC} \leq 4 \\ 0, & \text { otherwise }\end{cases}
\end{align}
where TTC is defined as $\mathrm{TTC}_i(t)=-\frac{s_i(t)}{v_{i-1}(t)-v_i(t)}$. We set the time-to-collision threshold as $4$ seconds, below which indicates dangerous driving and results in a negative reward. As demonstrated in \cite{das2019defining}, the statistical median of TTC for car-car leader-follower pairs is about $4$ seconds.

\subsection{Learning-Based Human Driver Behavior Identification}
\label{subsec:SI}
\begin{figure}[th]
    \centering
    \includegraphics[width=8.5cm]{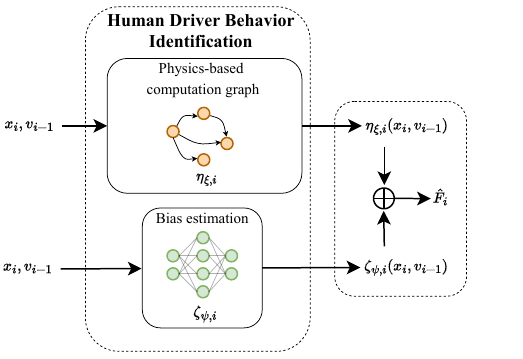}
    \caption{Learning-based human driver behavior identification module.}
    \label{fig: hdbi}
\end{figure}
Inspired by \cite{mo2021physics}, we employ a learning-based human driver behavior identification module to derive the unknown car-following parameters for each HDV as in Fig.~\ref{fig: hdbi}. The dynamics of the estimated car-following behavior for vehicle $i$, denoted by $ \hat{\mathbb{F}}_i(\cdot)$, is expressed as:
\begin{equation}
    \hat{\mathbb{F}}_i(s_i, v_i, v_{i-1}) = \alpha_1  s_i - \alpha_2  v_i + \alpha_3  v_{i-1}  + \zeta_{\psi,i}(s_i, v_i, v_{i-1}),
    \label{eq: learning-based system identification}
\end{equation}
where $\eta_{\xi,i}(s_i, v_i, v_{i-1}):= \alpha_1  s_i - \alpha_2  v_i + \alpha_3  v_{i-1}$ represents a physics-based computation graph parameterized by $\xi=[\alpha_1, \alpha_2, \alpha_3]$ to approximate the car-following behavior via a linearized model, and $\zeta_{\psi,i}$ is a neural network-based bias estimator parameterized by $\psi$ that quantifies the deviation between the actual nonlinear car-following dynamics and the estimations from computation graphs. 
The reason for dividing the car-following dynamics into a linear physics-based computation graph and a nonlinear bias estimator is that the construction of CBF requires the estimation of the partial derivatives of $\hat{\mathbb{F}}_i$, i.e., $\alpha_1=\frac{\hat{\mathbb{F}}_i}{\partial s_i}$, $\alpha_2=-\frac{\hat{\mathbb{{F}}_i}}{\partial v_{i}}$, $\alpha_3=\frac{\hat{\mathbb{F}}_i}{\partial v_{i-1}}$. Thereby, we separate the computation graph from the bias estimator to enable the  explicit training of $\xi=[\alpha_1, \alpha_2, \alpha_3]$ using real data since the form and parameters of $\hat{\mathbb{F}}_i(\cdot)$ are not known. 

It is worth noting that our system identification method is different from both (i) classical system identification considering only linear components and (ii) neural network-based car-following models. Unlike classical system identification, we estimate the nonlinear bias in the car-following model to enhance the precision of the estimation process. Unlike neural network-based car-following models that leverage a single neural network to predict acceleration rates, we combine a computation graph to explicitly estimate $\xi=[\alpha_1, \alpha_2, \alpha_3]$ to satisfy the requirement of CBF construction. 

Using the estimated system dynamics for each HDV $\hat{\mathbb{F}}_i$, we can formulate CBFs for both the following HDVs and the CAV.

\subsection{Differentiable Safety Module}
\label{subsec:QP}
In this subsection, we introduce the differential safety module that comprises two main components. The first component utilizes the CBF-QP-based framework to integrate safety guarantees into the DRL-based controller by converting the DRL actions to a safe one. To take the following vehicles' safety into account, this approach encompasses CBF-based safety constraints not only for the CAV but also for the following HDVs. Consequently, it goes beyond merely improving the safety of the ego vehicle but enhances the system-level safety of the platoon. The second component introduces a differentiable QP layer for neural networks that enables the backpropogation of the QP solution to parameters in the CBF-QP module and the DRL policy network such that both the safety layer and the DRL-based controller can be simultaneously trained. 
This module addresses the issue with \cite{xiao2023barriernet} that the safety filter does not provide guidance on the training of the DRL mode.

\subsubsection{CBF-QP-Based Approach to Incorporate Safety Guarantees} 
We first incorporate safety guarantees into DRL by converting the generated DRL  action to a safe action by solving a QP, whereby the safety constraints are characterized via CBF defined in Eq. \eqref{CBF definition}. To avoid collisions, it is essential for each vehicle to maintain a sufficient headway from the preceding vehicle. By adopting time headway as the safety metric, the safety conditions for all vehicles following the CAV can be expressed as:
\begin{equation}
    s_j\geq\tau v_j, j\in\left\{i,\cdots,n\right\} \label{eq:cbf_spacing}
\end{equation}
where $\tau$ is the minimum allowed time headway. CAV is indexed by $i$ and the following HDVs are indexed by $i+1,\cdots,n$. With Eq. (\ref{eq:cbf_spacing}), we have the following CBF candidate: 
\begin{equation}
    h_{\text{th},j}(x) =s_j-\tau v_j\geq 0, j\in\left\{i,\cdots,n\right\}
    \label{eq: original CBF candidate}
\end{equation}
We make the following remark for the CBF candidate Eq. (\ref{eq: original CBF candidate}), which can also be seen as a state constraint. 
\begin{remark}
Although alternative methods exist to handle Eq. (\ref{eq: original CBF candidate}), it is challenging for these methods to provide a formal safety guarantee in a computationally efficient manner. One common method is to integrate Eq. (\ref{eq: original CBF candidate}) into the DRL reward function, which, however, is only a soft constraint without a formal guarantee. Alternatively, Eq. (\ref{eq: original CBF candidate}) can be explicitly integrated into an optimization problem to serve as a state constraint, which can be inefficient to account for future states due to the myopic nature of the state constraint (i.e., it only involves a particular time step). In contrast, CBF as introduced in Def. \ref{cbf} can provide a theoretical safety guarantee in a computationally efficient manner via the property of \emph{forward invariance}, which ensures that if the current state is safe, there always exist control actions such that \emph{future states} remain safe.
\end{remark}

However, for the CBF candidate $h_{\text{th},j}(x)$, the relative degree\footnote{The relative degree represents the number of times we need to differentiate the CBF candidate along the dynamics of Eq.~\eqref{continuous system} until the control action $u$ explicitly shows.} is $j-i+1\geq 2$ for the $j$-th following vehicle, which cannot be handled by the original CBF. Although such an issue can be handled by high-order CBFs \cite{zhou2022safety}, obtaining the high-order derivatives of states in real-world systems is typically computationally expensive and imprecise. Therefore, we introduce an alternative approach to reduce the relative degrees. Specifically, instead of directly using Eq. (\ref{eq: CBF candidate}) as the CBF candidate, we utilize revised safety conditions:
\begin{equation}
    \begin{aligned}
        h_i(x) &=h_{\text{th},i}(x)=s_i-\tau v_i,\\
        h_j(x) &=h_{\text{th},j}(x)-h_{\text{th},i}(x)\\
        &=s_j-s_i-\tau (v_j - v_i), j\in\left\{i+1,\cdots,n\right\}\\
    \end{aligned}
    \label{eq: CBF candidate}
\end{equation}
where $h_i(x)$ is identical to $h_{\text{th},i}(x)$ to ensure that the headway between CAV and its preceding vehicle is sufficient. For $h_j(x), i =i,\cdots,n$, we tighten the safety conditions by requiring the difference between the CBF of HDV $j$ (i.e. $h_{\text{th},j}(x)$) and the CBF of CAV (i.e. $h_{\text{th},i}(x)$) to be nonnegative. Note that $h_j(x)\geq 0,~i=0,\cdots,n$ is a sufficient condition for $h_{\text{th},j}(x)\geq 0,~j=i+1,\cdots,n$, which ensures system-level safety. Moreover, we notice that the relative degree of $h_j(x)$ remains $1$, which is convenient for the calculation and real-world implementation.

Next, we compute the derivatives of $h_i(x)$ and $h_j(x)$ as $\nabla h_i(x) = \left[1,-\tau,\cdots,0,0\right]^\top$ and $\nabla h_j(x)=\left[-1, \tau,0,0,\cdots,1,-\tau,\cdots\right]^\top$. The Lie derivatives of the CBF candidates in Eq. \eqref{eq: CBF candidate} are given as:
\begin{subequations}
    \begin{align}
        &L_{f}h_i=\nabla h_i(x)^Tf(x) = v_{i-1}-v_i,\\
        &L_{f}h_j=\nabla h_j(x)^Tf(x) = v_i - v_{i-1} - v_j + v_{j-1} \notag \\
        & \quad\quad -\tau\hat{\mathbb{F}}_j\left(x_j,v_{j-1}\right), j\in\left\{i+1,\cdots,n\right\}, \\
        &L_gh_i = \nabla h_i(x)^Tg(x) = -\tau,\\
&L_gh_j = \nabla h_j(x)^Tg(x) = \tau, j\in\left\{i+1,\cdots,n\right\}
    \end{align}
\end{subequations}

Using the CBF definition~\eqref{CBF definition} and Lie derivatives of the CBF candidate calculated above, the CBF constraint is given by:
\begin{subequations}
    \begin{align}
        &L_{f}h_i + L_{g}h_iu + \alpha_{\text{CBF},i}(h_i) \geq 0,\\
        &L_{f}h_j + L_{g}h_ju + \alpha_{\text{CBF},j}(h_j) +\sigma_j \geq 0, j\in\left\{i+1,\cdots,n\right\}
    \end{align}
\label{eq:cbf_lcc}
\end{subequations}

\noindent where the class $\mathcal{K}_{\infty}$ function $\alpha_{\text{CBF},i}(\cdot), \cdots, \alpha_{\text{CBF},n}(\cdot)$ are set to linear functions with positive coefficients $k_i,\cdots,k_n$ serving as the parameters to be trained. The choice of a linear class $\mathcal{K}_{\infty}$ function follows the insights of \cite{xiao2021high}, which showed that using a linear class $\mathcal{K}_{\infty}$ function for CBFs, compared to a quadratic one, results in smaller, more gradual control inputs that can respect CAV actuation limits and enhance passenger comfort. $\sigma_{i+1} \cdots \sigma_n$ are the slack variables for HDV safety constraints to avoid conflict with CAV safety constraints. To better tailor the derived CBF constraint to the DRL settings, we rewrite Eq.~\eqref{eq:cbf_lcc} as follows: 
\begin{subequations}
    \begin{align}
    &L_{f}h_i + L_{g}h_i(u_{\text{safe}} + u_{\text{RL}}) + k_ih_i \geq 0,\label{eq: CAV safety constraint}\\
    &L_{f}h_j + L_{g}h_j(u_{\text{safe}} + u_{\text{RL}}) + k_jh_j+\sigma_j \geq 0, j\in\left\{i+1,\cdots,n\right\}
    \end{align}
\end{subequations}
\noindent where $u=u_{\text{RL}}+u_{\text{safe}}$ with $u_{\text{RL}}$ representing the action provided by the DRL algorithm. We divide the control input into $u_{\text{RL}}$ and $u_{\text{safe}}$ to highlight the compensation effect of the CBF-QP. Notice that directly optimizing $u$ is equivalent to optimizing $u_{\text{safe}}$, as $u_{\text{RL}}$ is treated as given in the QP problem.

Moreover, due to the actuator limitations, we have the acceleration constraint as follows:
\begin{equation}
    a_{\min}\leq u_{\text{RL}}+u_{\text{safe}}\leq a_{\max}, \label{QP_con2}
\end{equation}
where $a_{\max}$ and $a_{\min}$ denote the maximum and minimum acceleration rates for CAVs, respectively. However, the acceleration constraint may conflict with the CAV safety constraint~Eq. (\ref{eq: CAV safety constraint}) and make the QP problem infeasible, which is because the CAV safety constraint is a hard constraint without a relaxation term. To guarantee the feasibility of the QP problem, we integrate a feasibility constraint as in \cite{xiao2022sufficient,xu2022feasibility}. By utilizing this approach, we enforce a sufficient condition for the feasibility via another CBF in the mixed-autonomy platoon. The feasibility constraint is formulated as follows:
\begin{equation}
    u_{\text{safe}}+u_{\text{RL}} \leq \hat{\mathbb{F}}_{i-1}+k_{\text{f}}(v_{i-1}-v_i-\tau a_{\min }),
    \label{eq:feasibility constraint}
\end{equation}
with a positive trainable parameter $k_{\text{f}}$ and estimated acceleration of the preceding vehicle $\hat{\mathbb{F}}_{i-1}$. The feasibility constraint provides a conflict-free guarantee of the QP problem corresponding to the next time interval, the details of which can be found in \cite{xu2022feasibility}.

Next, we design the CBF-QP controller for the CAV. The objective is to minimize the control input deviation $u_{\text{safe}}$ and relaxation term $\sigma_{i+1},\cdots\,\sigma_n$ under the above-mentioned constraints. To this end, we form the QP optimization problem yielding
\begin{equation}
    \begin{aligned}
   &\min_{u_{\text{safe}},\sigma_{i+1},\cdots,\sigma_n} \quad  ||u_{\text{safe}}||^{2}+ \sum_{j=i+1}^n b_j\sigma_j^2, \\ 
   \text {s.t.}  \quad  &L_{f}h_i + L_{g}h_i(u_{\text{safe}} + u_{\text{RL}}) + k_ih_i \geq 0,\\
        &L_{\hat{\mathbb{F}}_j}h_j + L_{g}h_j(u_{\text{safe}} + u_{\text{RL}}) + k_jh_j +\sigma_j \geq 0, \\
        & \quad \quad  j\in\{i+1,\cdots,n\}\\
        &u_{\text{safe}} + u_{\text{RL}} \leq \hat{\mathbb{F}}_{i-1}+k_{\text{f}}(v_{i-1}-v_i-\tau a_{\min }),\\
        & a_{\min} \leq u_{\text{safe}} + u_{\text{RL}} \leq a_{\max},
    \end{aligned}\label{eq: control optimization}
\end{equation}
where $b_j>0,~j=i+1,\cdots,n$ is the penalty coefficient for the slack variable $\sigma_j$. Here,  $b_j>0,~j=i+1,\cdots,n$ are all set to $1$ for simplicity. Parameters $ \theta_{\text{CBF}} = \left\{\{k_j\}_{j=i}^n, k_\text{f}\right\}$ can be trained simultaneously with the RL policy network via a differentiable QP layer, which allows the CBF constraints to better adapt to the specific environment, as presented next. 

\subsubsection{Differentiable QP for Neural Networks}
\label{sec:safety}
Then, we present a differentiable QP to enable the backpropagation of the QP solution derived in Section~\ref{subsec:QP} to facilitate the training of RL. 

Let us denote the loss function by $\ell$, and the optimal solution of the CBF-QP is $w^{\star}= \left(u_{\text{safe}}, \sigma_1,\cdots,\sigma_n\right)$.  
Let $\theta=\left\{\theta_{\text{RL}},\theta_{\text{CBF}}\right\}$ represent the trainable parameters in the DRL actor network and the CBF-QP controller. 
To train these parameters, we are interested in calculating the following partial derivatives, which are obtained based on the chain rule: 
\begin{align}
    \frac{\partial \ell}{\partial \theta_{\text{RL}}} &= 
    \frac{\partial \ell}{\partial u}\frac{\partial u}{\partial \theta_{\text{RL}}} = \frac{\partial \ell}{\partial u}\Big(\frac{\partial u_{\text{RL}}}{\partial \theta_{\text{RL}}} + \frac{\partial u_{\text{safe}}}{\partial \theta_{\text{RL}}} \Big) \notag \\ 
    & = \frac{\partial \ell}{\partial u}\Big(\frac{\partial u_{\text{RL}}}{\partial \theta_{\text{RL}}} + \frac{\partial u_{\text{safe}}}{\partial u_{\text{RL}}}\frac{\partial u_{\text{RL}}}{\partial \theta_{\text{RL}}} + \frac{\partial u_{\text{safe}}}{\partial \theta_{\text{CBF}}}\frac{\partial \theta_{\text{CBF}}}{\partial \theta_{\text{RL}}}\Big) \notag \\     
    & = \frac{\partial \ell}{\partial u}\Big(\frac{\partial u_{\text{RL}}}{\partial \theta_{\text{RL}}} + \frac{\partial u_{\text{safe}}}{\partial u_{\text{RL}}}\frac{\partial u_{\text{RL}}}{\partial \theta_{\text{RL}}}\Big) , \label{eq:partial_RL}\\
    \frac{\partial \ell}{\partial \theta_{\text{CBF}}} &= 
    \frac{\partial \ell}{\partial u}\frac{\partial u}{\partial \theta_{\text{CBF}}} = \frac{\partial \ell}{\partial u}\left(\frac{\partial u_{\text{RL}}}{\partial \theta_{\text{CBF}}} + \frac{\partial u_{\text{safe}}}{\partial \theta_{\text{CBF}}}\right) = \frac{\partial \ell}{\partial u} \frac{\partial u_{\text{safe}}}{\partial \theta_{\text{CBF}}} \label{eq:partial_CBF}
\end{align}
where the first equality condition in both Eq. (\ref{eq:partial_RL}) and Eq. (\ref{eq:partial_CBF}) results from the chain rule, and the second equality condition in both equations results from the relation $u = u_{\text{RL}} + u_{\text{safe}}$. The third equality condition in Eq. (\ref{eq:partial_RL}) is because by Eq. (\ref{eq: control optimization}), $u_{\text{safe}}$ can be seen as a function with respect to $u_{\text{RL}}$ and $\theta_{\text{CBF}}$. The last equality condition of Eq. (\ref{eq:partial_RL}) is due to the relation $\frac{\partial \theta_{\text{CBF}}}{\partial \theta_{\text{RL}}} = 0$ because the CBF-QP parameters $\theta_{\text{CBF}}$ do not depend on the actor network. The last equality condition of Eq. (\ref{eq:partial_CBF}) is due to the relation $\frac{\partial u_{\text{RL}}}{\partial \theta_{\text{CBF}}} =0$ because the actor network does not depend on the parameters $\theta_{\text{CBF}}$.
Note that $\frac{\partial \ell}{\partial u}$ can be easily obtained from the loss function, and $\frac{\partial u_{\text{RL}}}{\partial \theta}$ is defined by the architecture of the actor network. Therefore, we only need to calculate $\frac{\partial u_{\text{safe}}}{\partial u_{\text{RL}}}$ and $\frac{\partial u_{\text{safe}}}{\partial \theta_{\text{CBF}}}$, which are part of $\frac{\partial w^{\star}}{\partial u_{\text{RL}}}$  and $\frac{\partial w^{\star}}{\partial \theta_{\text{CBF}}}$ as presented next.

Let us rewrite the aforementioned CBF-QP in Eq.~\eqref{eq: control optimization} in a more general form with decision variable $w=\left(u_{\text{safe}}, \sigma_1,\cdots,\sigma_n\right)$: 
\begin{equation}
    \begin{aligned}
    \min_w \quad & \frac{1}{2} w^T Q w \\
    \text { subject to } \quad & G w \leq q(u_{\text{RL}}, \theta_{\text{CBF}})
    \end{aligned}
    \label{safe QL}
\end{equation}
where $Q, G, q$ are the corresponding parameters of the QP problem. Note that only $q$ is a function of $u_{\text{RL}}$ and $\theta_{\text{CBF}}$. As in~\cite{xiao2023barriernet}, the calculation of $\frac{\partial w^{\star}}{\partial u_{\text{RL}}}$ and $\frac{\partial w^{\star}}{\partial \theta_{\text{CBF}}}$ can be performed by differentiating the KKT conditions with equality, i.e.,  
\begin{equation}
\begin{aligned}
Q w^{\star}+G^T \lambda^{\star} & =0, \\
\mathbb{D}\left(\lambda^{\star}\right)\left(G w^{\star}-q(u_{\text{RL}}, \theta_{\text{CBF}})\right) & =0,
\end{aligned} \label{eq:KKT}
\end{equation}
where $w^{\star}$ and $\lambda^{\star}$ represent the optimal primal and dual variables, and $\mathbb{D}(\lambda^\star)$ is a diagonal matrix constructed from vector $\lambda^\star$. 

Since the calculation of $\frac{\partial u_{\text{safe}}}{\partial u_{\text{RL}}}$ and $\frac{\partial u_{\text{safe}}}{\partial \theta_{\text{CBF}}}$ is similar, we next only illustrate the calculation of $\frac{\partial u_{\text{safe}}}{\partial u_{\text{RL}}}$. For the calcluation of $\frac{\partial w^{\star}}{\partial \theta_{\text{CBF}}}$, we only need to replace $u_{\text{RL}}$ with $\theta_{\text{CBF}}$ in Eq.\eqref{derivative of KKT}. 
Then we take the derivative of Eq.\eqref{eq:KKT} with respect to the RL action $u_{\text{RL}}$ and write in the matrix form as:
\begin{equation}
\begin{aligned}
&  {\left[\begin{array}{l}
\displaystyle \frac{\partial w^{\star}}{\partial u_{\text{RL}}} \\
 \displaystyle \frac{\partial  \lambda^{\star}}{\partial u_{\text{RL}}} 
\end{array}\right]=}  K^{-1}\left[\begin{array}{c}
\mathbb{O} \\
\mathbb{D}\left(\lambda^{\star}\right)\displaystyle \frac{\partial  q(u_{\text{RL}},\theta_{\text{CBF}})}{\partial u_{\text{RL}}} 
\end{array}\right],
\end{aligned}
\label{derivative of KKT}
\end{equation}
with $K = \left[\begin{array}{cc}
Q & G^T \\
\mathbb{D}\left(\lambda^{\star}\right) G & \mathbb{D}\left(G w^\star-q(u_{\text{RL}},\theta_{\text{CBF}})\right)  
\end{array}\right]$. 

Utilizing the partial derivative of the QP layer, the parameters of both DRL and CBF constraints can be simultaneously updated, allowing for adaptive adjustments to the training environment.

\section{Simulation Results}
\label{Simulation Results}
In this section, we evaluate the proposed method in simulations of a typical mixed-autonomy platoon. We consider a typical platooning scenario as depicted in Fig.~\ref{fig:CAV_overview}, which features a platoon of five vehicles traveling on a single lane, with the third vehicle being a CAV and the others HDVs. For benchmarks, we compare four types of models: pure car following, PPO without the safety layer (PPO without safety guarantee), PPO with the safety layer but without learning-based human driver behavior identification (i.e., safe-RL without SI), and PPO with the safety layer and learning-based human driver behavior identification (i.e., safe-RL with SI).

\subsection{Training Settings and Results}
\begin{figure}[t]
    \centering
    \includegraphics[width=6cm]{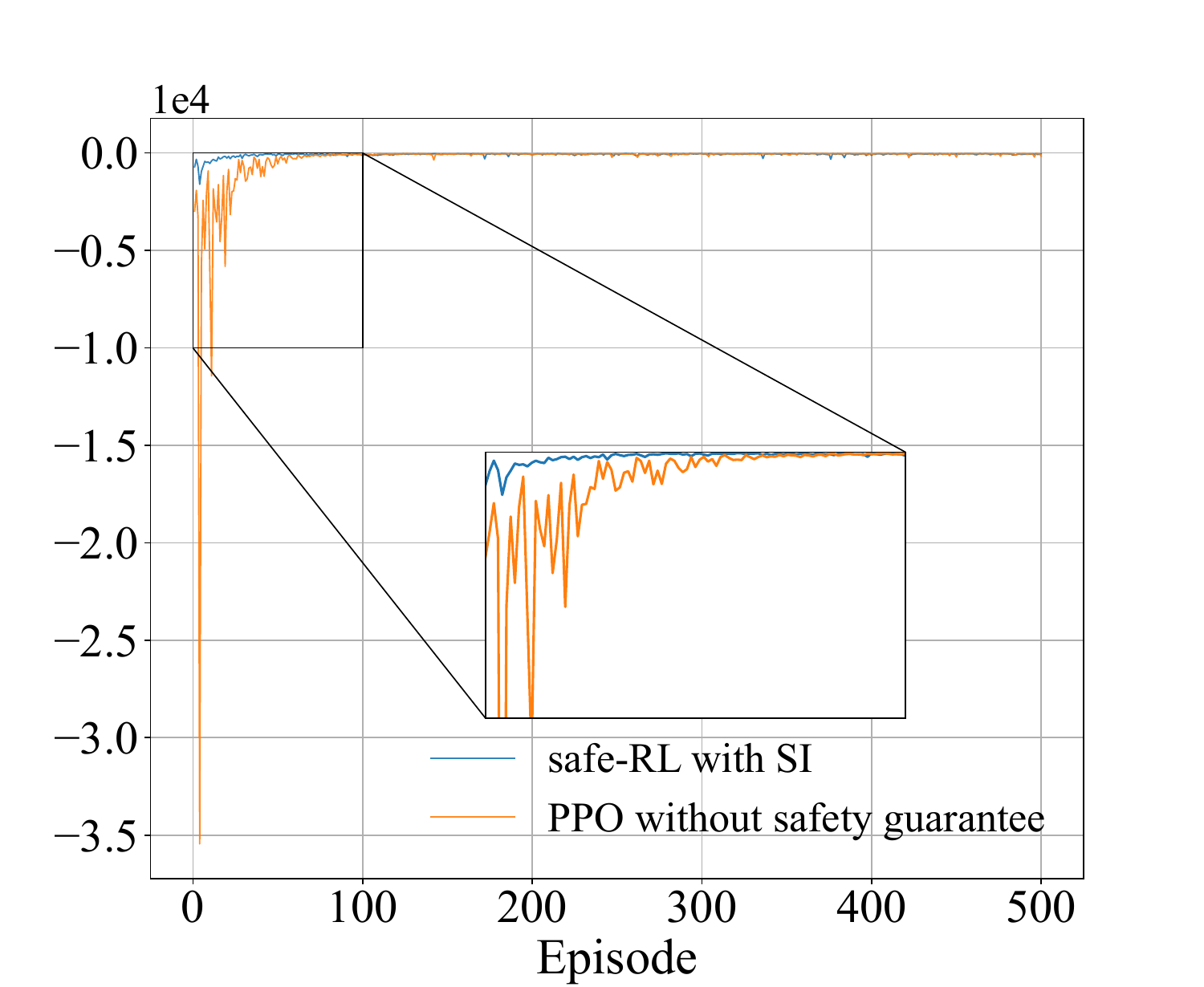}
    \caption{Training rewards per episode.}
    \label{fig:training}
\end{figure}
In the training scenario, we utilize a random velocity disturbance setup, wherein the head vehicle's velocity disturbance is sampled from a Gaussian distribution with zero mean and a standard deviation of $2$\,m/s independently at each time step within an episode. This chosen velocity disturbance is then multiplied by the time step ($0.1$\,s) and added to the head vehicle's original velocity. 

Without loss of generality, we choose the optimal velocity model (OVM) \cite{bando1998analysis} as the car-following model $\mathbb{F}$, following the setting of LCC~\cite{wang2021leading}. Note that the OVM can be replaced by any car-following model without influencing the applicability of the proposed method. The OVM reads as follows:
\begin{equation}
    \mathbb{F}(\cdot)=\alpha\left(V\left(s_i(t)\right)-v_i(t)\right)+\beta (v_{i-1}(t)-v_{i}(t)),
\end{equation}
where constants $\alpha,\beta>0$ represent car-following gains.
$V(s)$ denotes the spacing-dependent desired velocity of HDVs with the form given in Eq. (\ref{eq:optimal_speed}), where $s_{\mathrm{st}}$ and $s_{\mathrm{go}}$ represent the spacing thresholds for stopped and free-flow states, respectively, and  $v_{\mathrm{max}}$ denotes the free-flow velocity. 
\begin{equation}
V(s)= 
\begin{cases}
0, & s \leq s_{\mathrm{st}} \\
\displaystyle\frac{v_{\max }}{2}\left(1-\cos \left(\pi \frac{s-s_{\mathrm{st}}}{s_{\mathrm{go}}-s_{\mathrm{st}}}\right)\right), & s_{\mathrm{st}}<s<s_{\mathrm{go}} \\ 
v_{\mathrm{max}}, & s \geq s_{\mathrm{go}}
\end{cases} \label{eq:optimal_speed}
\end{equation}
The parameters of OVM is set as $\alpha = 0.6,\beta=0.9,s_{\text{st}}=5,s_{\text{go}}=35$. The equilibrium spacing and velocity are $20$ m and $15$ m/s.

The training hyperparameters are as follows. The learning rates for the actor and critic networks are both set at 0.0003 and we employ a linear decay schedule for the learning rate to facilitate training. As such, the performance of the algorithms is not sensitive to the initial learning rate and we choose the ones that can make the training process converge rapidly. The batch size is $2048$, the decay factor of the advantage function is $0.95$, the PPO clip parameter is set to $0.2$, and the number of experience utilizing is set to $10$. The training episodes are $500$. For the CBF parameters, the minimum time headway $\tau$ is set to $0.3$s and the initial values for the trainable parameters $\left\{k_0,k_1,k_2,k_f\right\}$ are $\left\{1,1,1,10\right\}$.

Under the specified training scenarios and parameter settings, the training rewards per episode for PPO without safety guarantee (i.e., without safety layer) and safe-RL with SI (i.e., with safety layer) are depicted in Fig.~\ref{fig:training}. Both algorithms converge after a sufficient number of training episodes, indicating that the designed reward function is suitable for mixed-autonomy platoon environments. Notably, the algorithms with a safety layer converge faster. This is expected as the safety layer reduces the range of feasible parameters and thus can accelerate training, especially in scenarios with large parameter space.

\begin{figure}
    \centering
    \includegraphics[width=7cm]{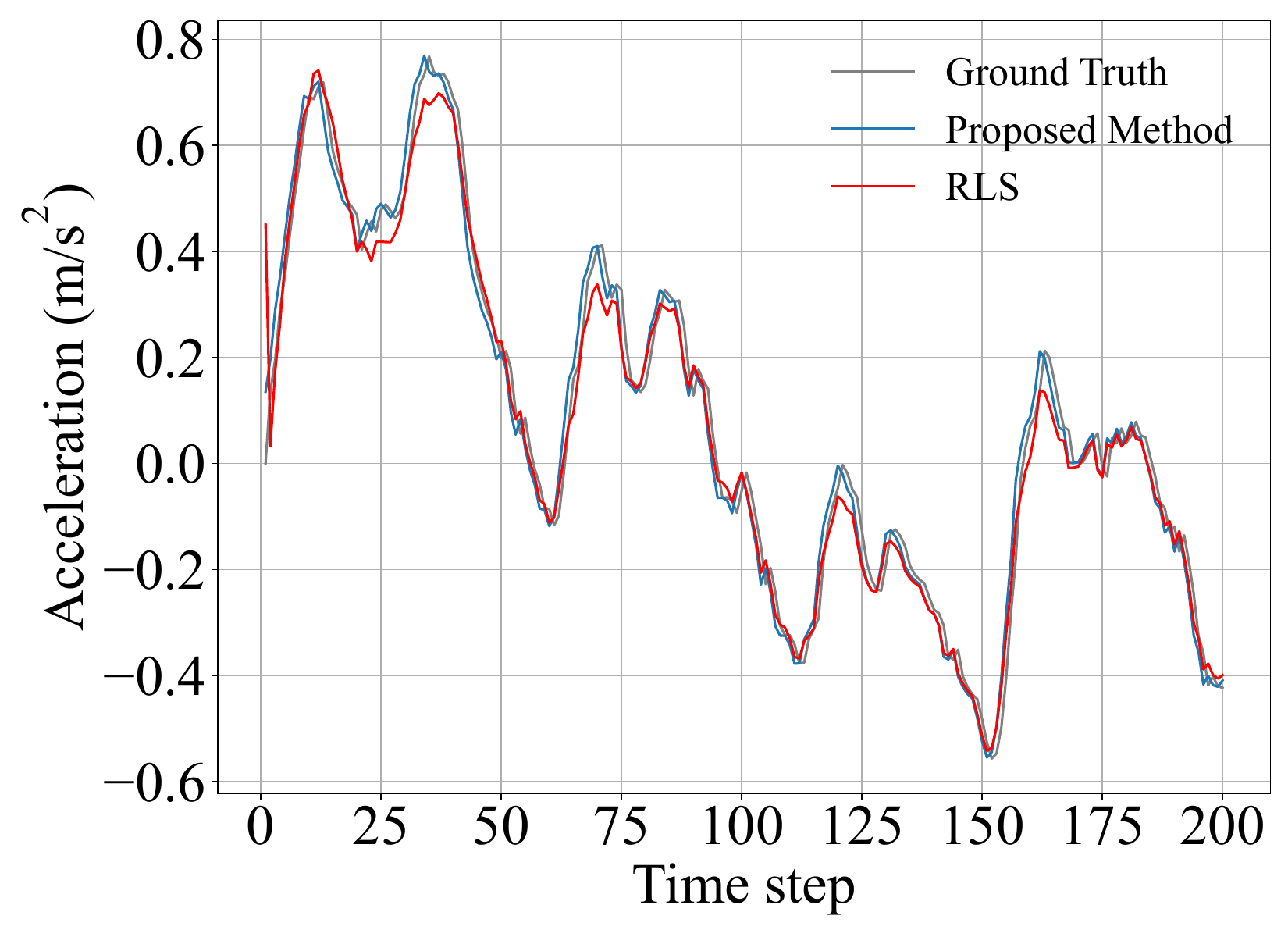}
    \caption{Comparison between online learning-based system identification and RLS.}
    \label{fig: SI comparison}
\end{figure}

For the learning-based driver behavior identification module as in Eq. \eqref{eq: learning-based system identification}, we train the parameters $\eta_{\xi,i}$ for the computation graph and $\zeta_{\psi,i}$ for the neural network-based bias estimator. The learning rate is set to 0.0001. To illustrate the benefits of our proposed learning-based driver behavior identification method, we conduct a comparison with recursive least squares (RLS) in terms of the estimation accuracy of vehicle $4$. The mean square error is 0.0019 for the proposed method and 0.0024 for RLS. The comparison results are given in Fig. \ref{fig: SI comparison}, which shows that our proposed method yields better accuracy for acceleration rate prediction compared with RLS. 

\subsection{Testing Results}
In the testing phase, two safety-critical scenarios that could cause safety-critical failures for the LCC model are simulated. Scenario 1 considers the scenario where the preceding HDV may brake urgently to avoid collisions with some unexpected cutting-in vehicle or crossing pedestrians, which may cause the distance to the CAV behind to be less than the safety distance. Scenario 2 describes a situation where the HDVs following the CAV in the platoon may suddenly accelerate due to human mistakes caused by driving fatigue. Such an instantaneous acceleration could endanger the vehicles ahead of it. Note that both scenarios are significantly different from the training scenarios with random disturbances. Such a treatment is to demonstrate the performance of the safety layer under adversarial conditions and the generalizability of our model to unseen scenarios.  For each of these two scenarios, we perform (i) a safety-guaranteed region analysis to quantitatively analyze the system-level safety improvement resulting from our proposed method, considering various acceleration rates and acceleration durations, and (ii) a case study to analyze the safety performance under a specific acceleration rate and acceleration duration for the two safety-critical scenarios. 

\subsubsection{Safety-Guaranteed Region Analysis}
\begin{figure}[ht]
    \centering
    \subcaptionbox{}{
            \includegraphics[width=4cm]{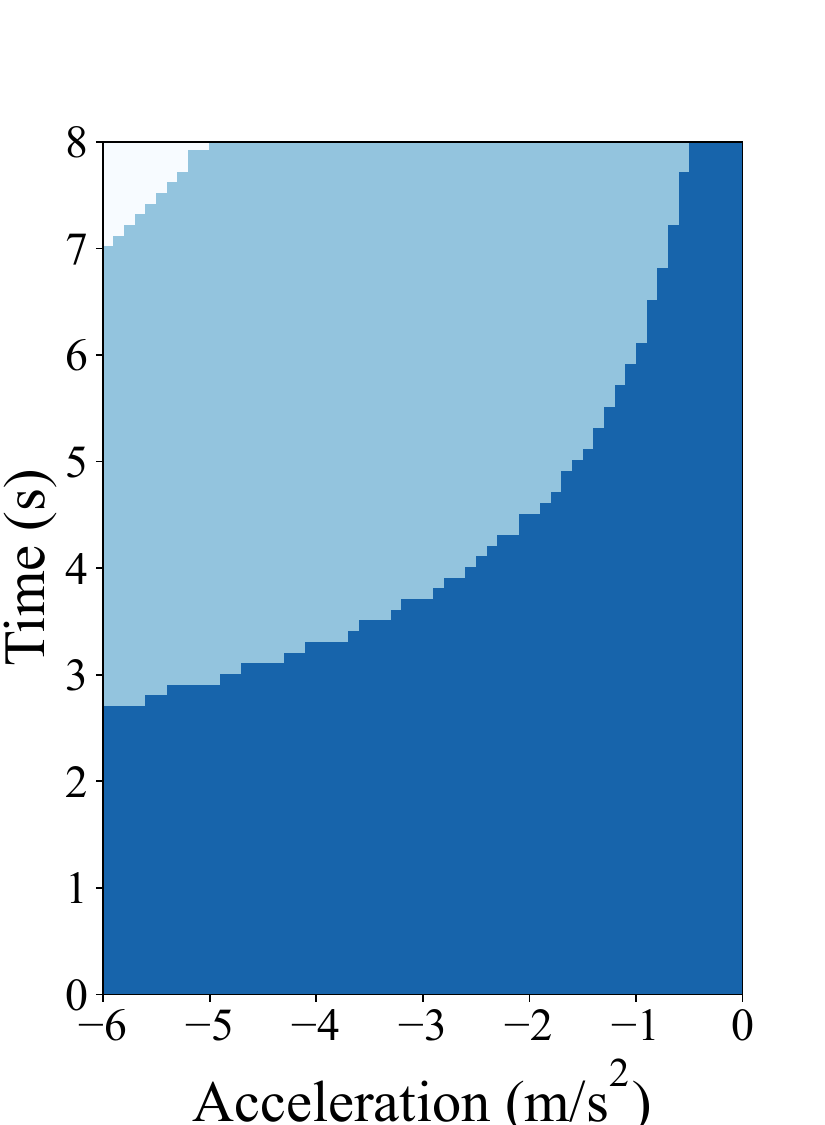}
        }
    \subcaptionbox{}{
            \includegraphics[width=4cm]{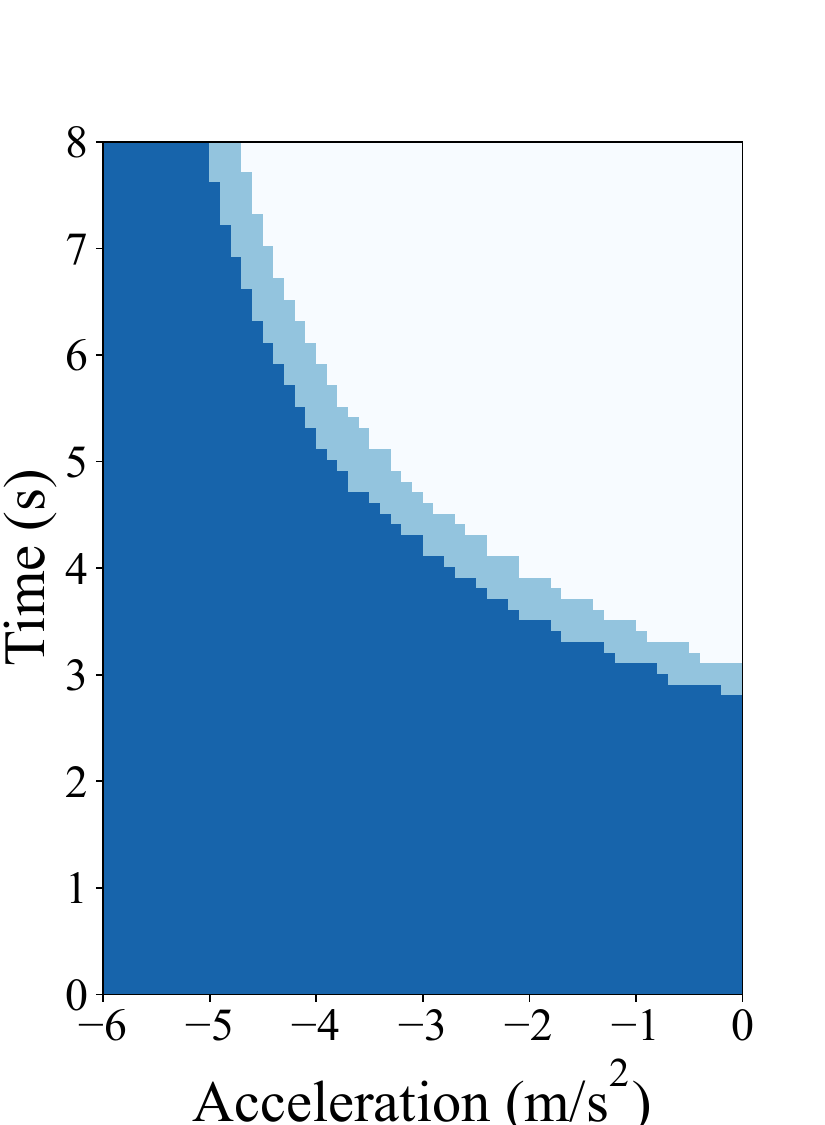}
        }
    \caption{Safety-guaranteed regions associated with two specific scenarios: (a) the deceleration disturbance of the preceding vehicle and (b) the acceleration disturbance of the following vehicle. The horizontal axis denotes the magnitude of the disturbance signal (acceleration/deceleration), and the vertical axis is the duration of the disturbance signal. The dark blue regions denote the safety region of the LCC with PPO controller without the safety layer, and the light blue regions illustrate the expanded safety region achieved by implementing the safe RL controller for LCC (i.e., with safety layer), while the white regions indicate the unsafe region.}
    \label{fig: safety region}
\end{figure}
To demonstrate the safety improvement of the platoon due to the incorporation of the safety layer, we show the resulting safety region with and without the safety layer in both safety-critical scenarios. In Fig.~\ref{fig: safety region} (a), the HDV indexed in $1$ decelerates for a range of duration times and deceleration rates. It can be seen that the safety region is expanded by almost 60\% using the proposed safe RL controller. Moreover, in Fig.~\ref{fig: safety region} (b), the HDV with index $3$ accelerates within a range of duration times and acceleration rates. Utilizing the proposed method, the following HDV's safety duration time increases by an average of 0.8s. In summary, our method effectively improves the safety regions in both scenarios.

\subsubsection{Case Study for Scenario 1}
\begin{figure}[ht]
    \centering
    \includegraphics[width=7cm]{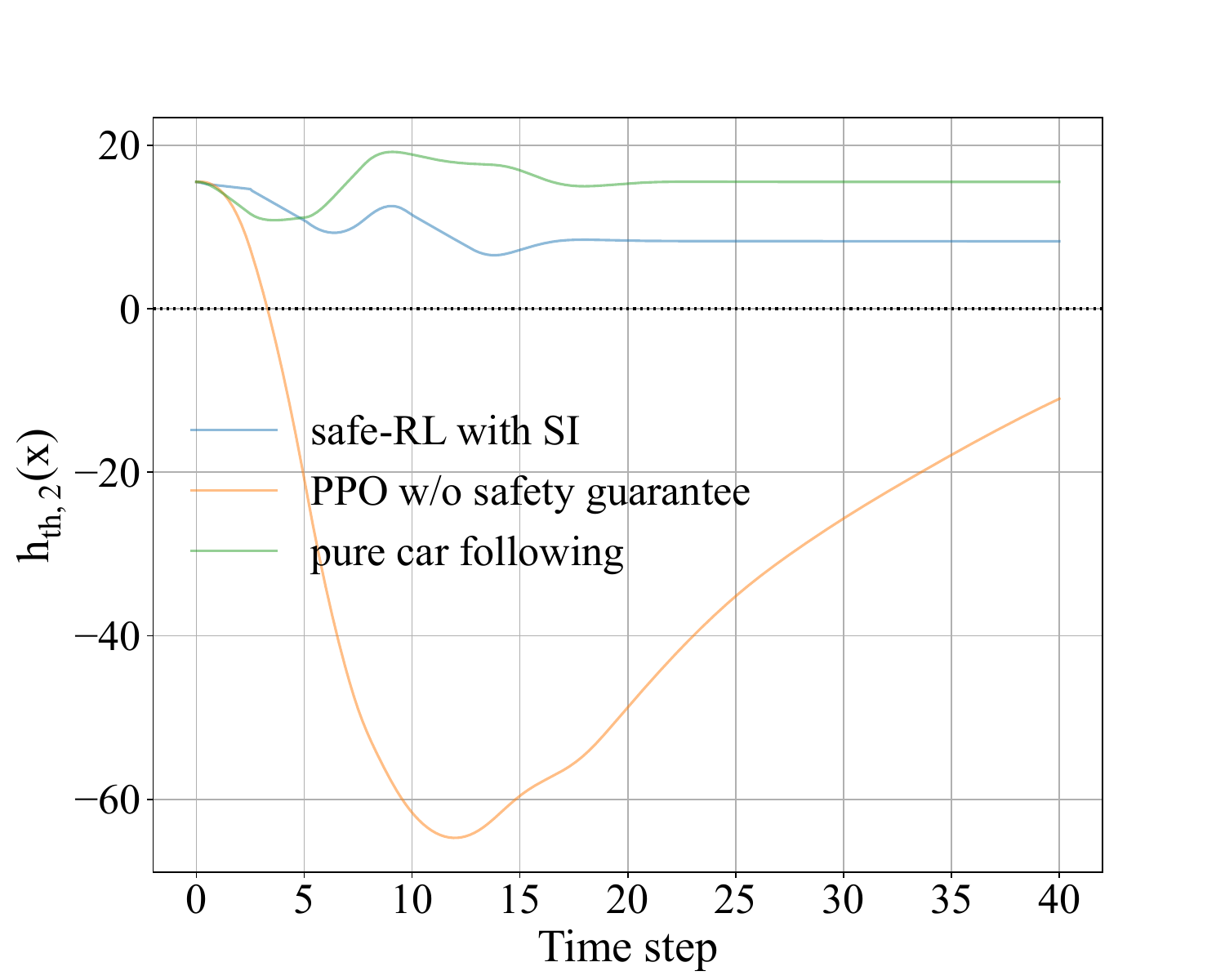}
    \caption{Value of the CBF candidate for CAV when the preceding vehicle emergency decelerates.}
    \label{fig: CAV safety}
\end{figure}
\begin{figure*}[ht]
    \centering
    \subcaptionbox{Spacing}{
            \includegraphics[width=5cm]{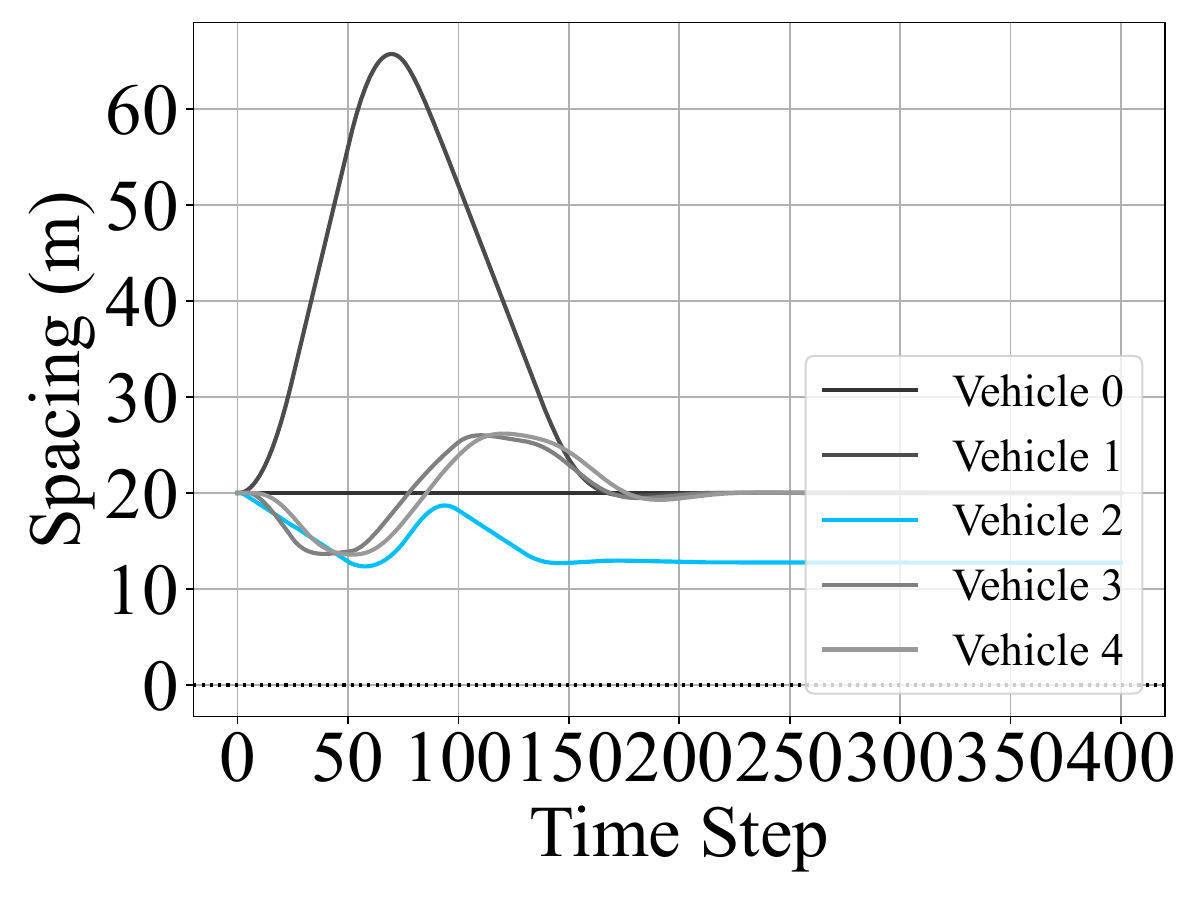}
        }
    \subcaptionbox{Velocity}{
            \includegraphics[width=5cm]{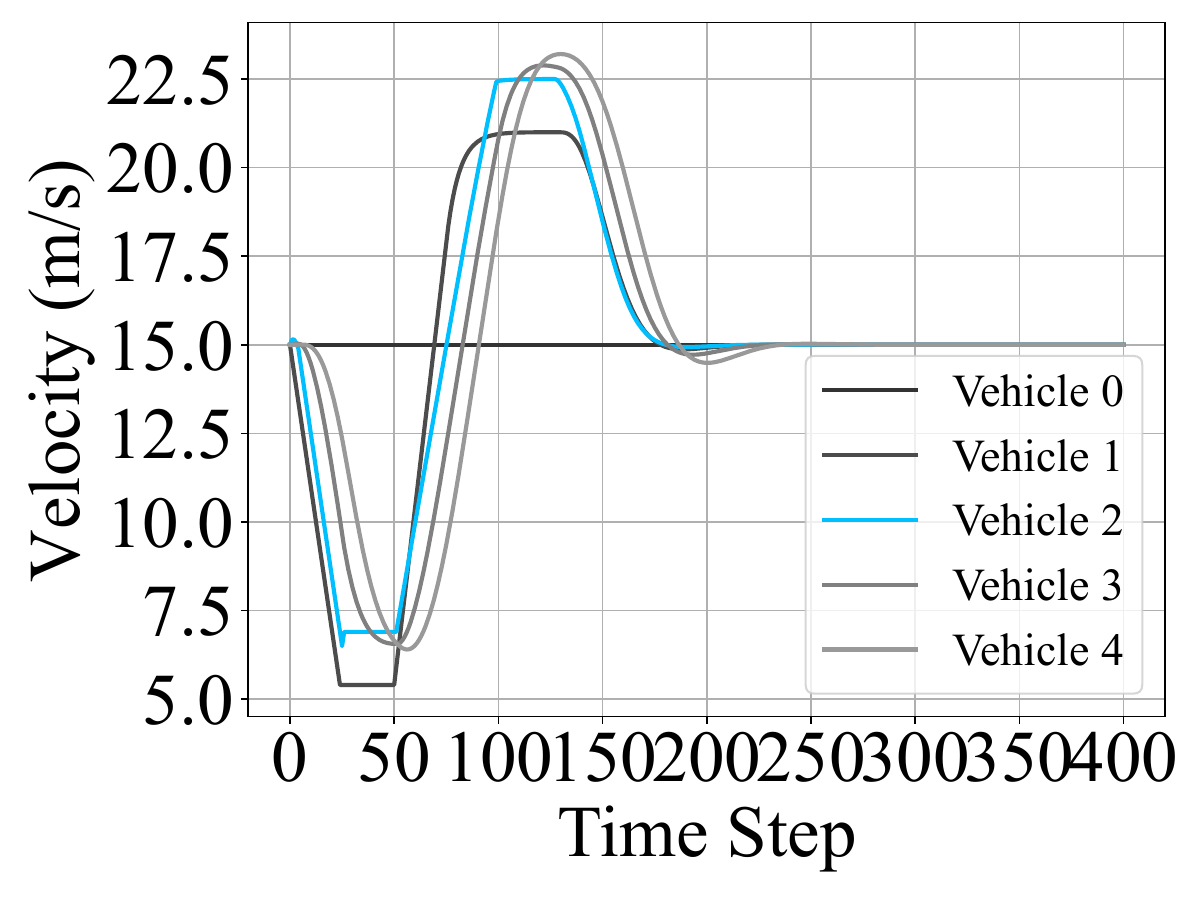}
        }
    \subcaptionbox{Acceleration}{
            \includegraphics[width=5cm]{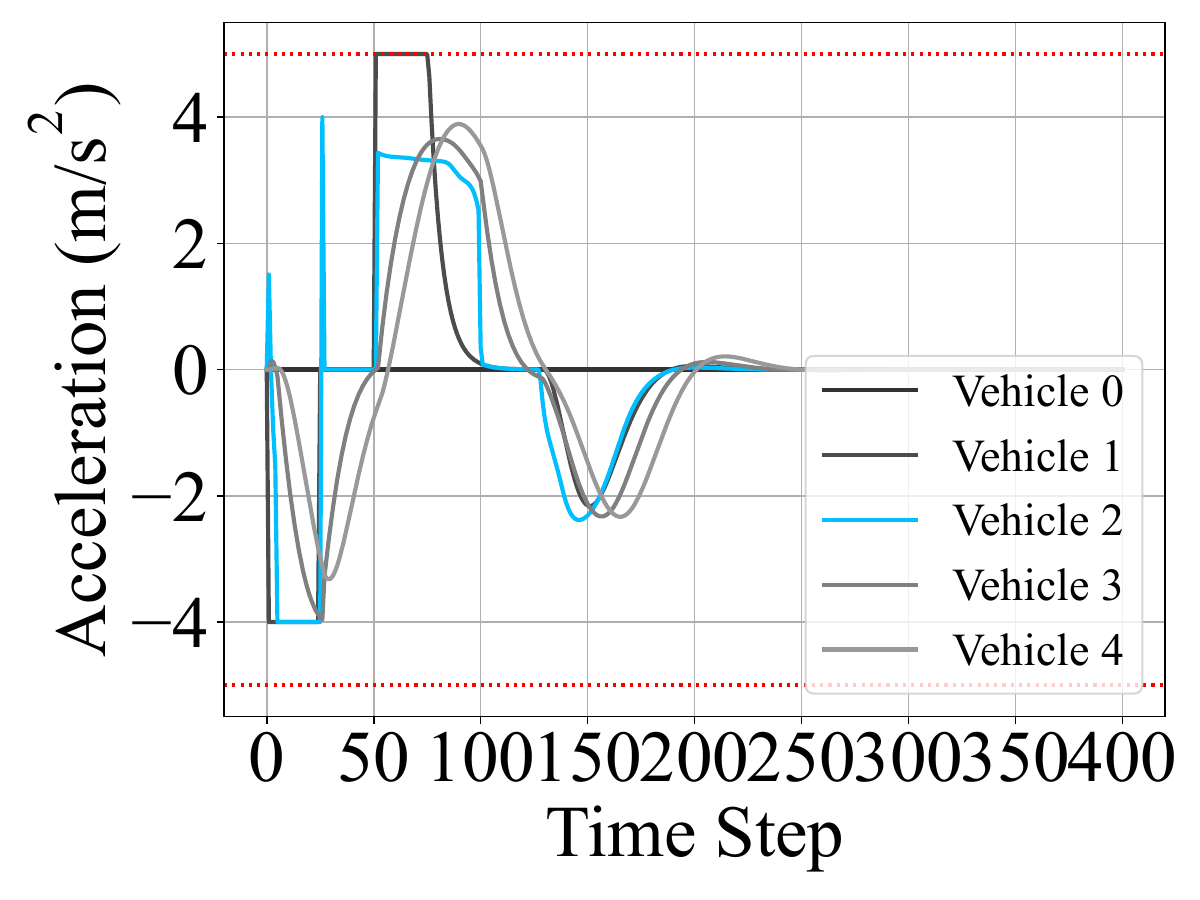}
        }
    \subcaptionbox{Spacing}{
            \includegraphics[width=5cm]{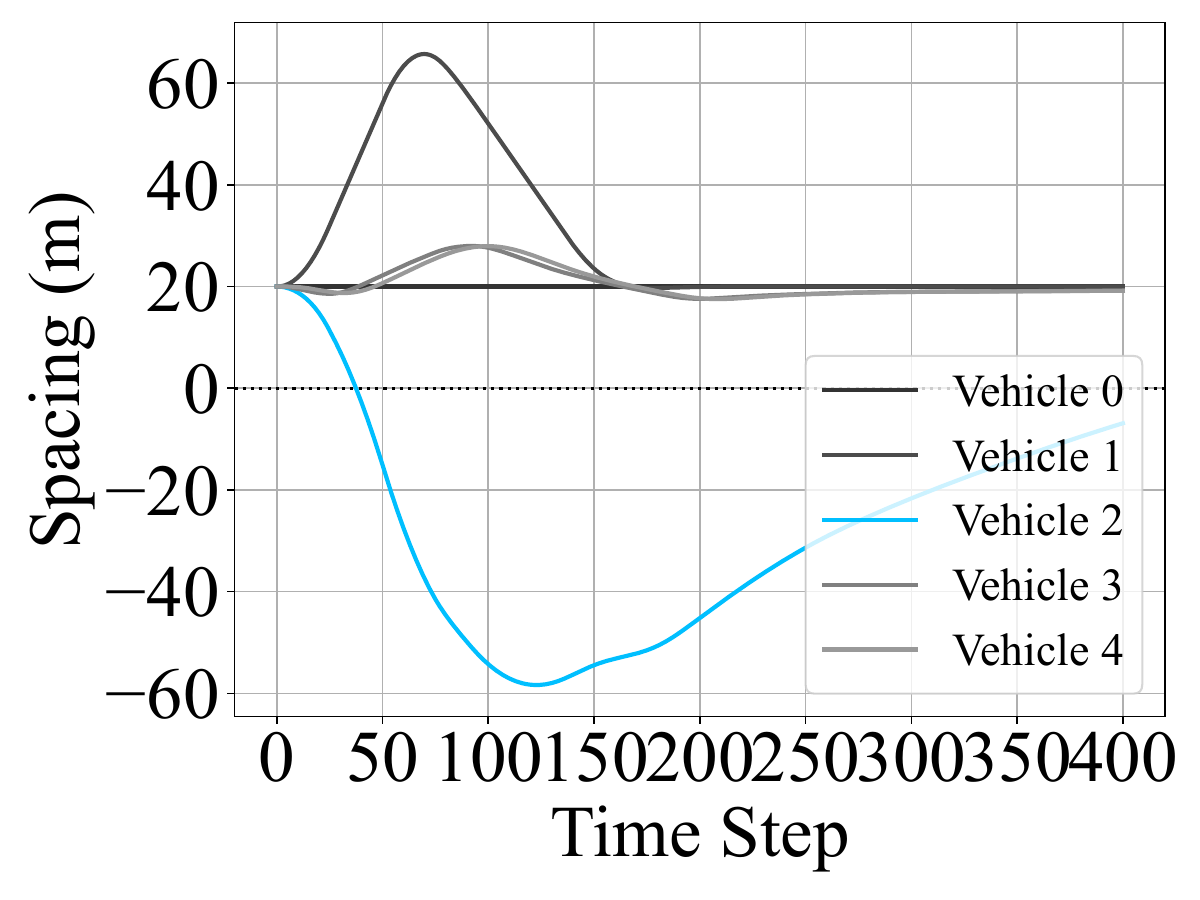}
        }
    \subcaptionbox{Velocity}{
            \includegraphics[width=5cm]{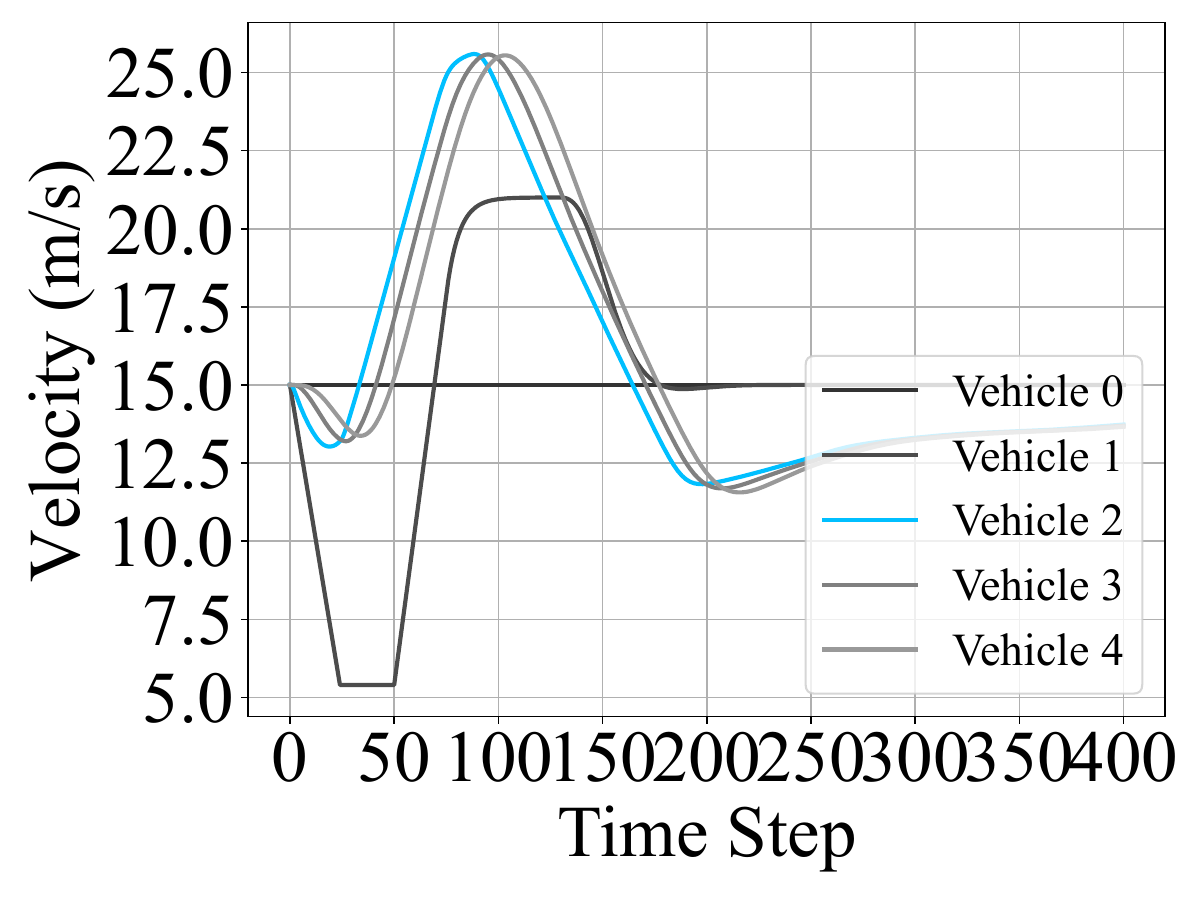}
        }
    \subcaptionbox{Acceleration}{
            \includegraphics[width=5cm]{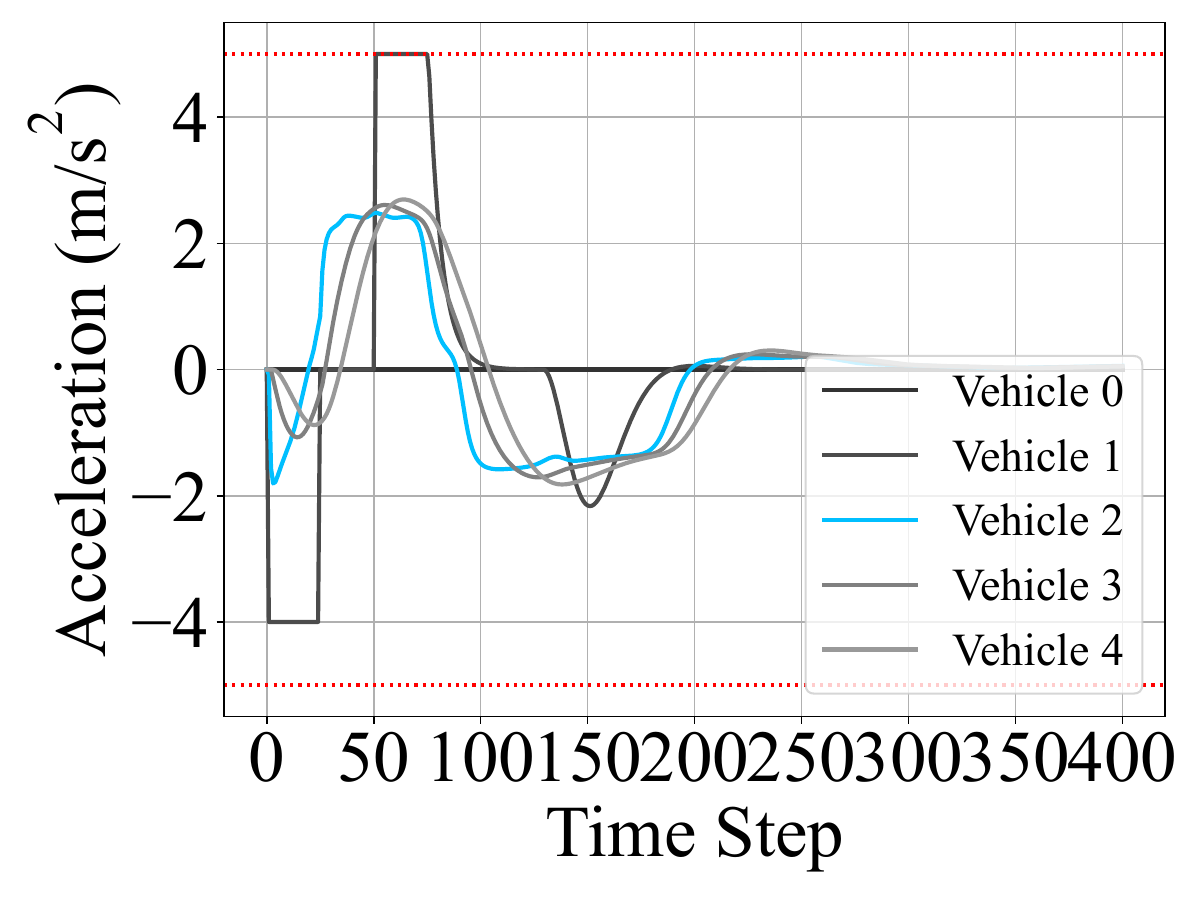}
        }
    \subcaptionbox{Spacing}{
            \includegraphics[width=5cm]{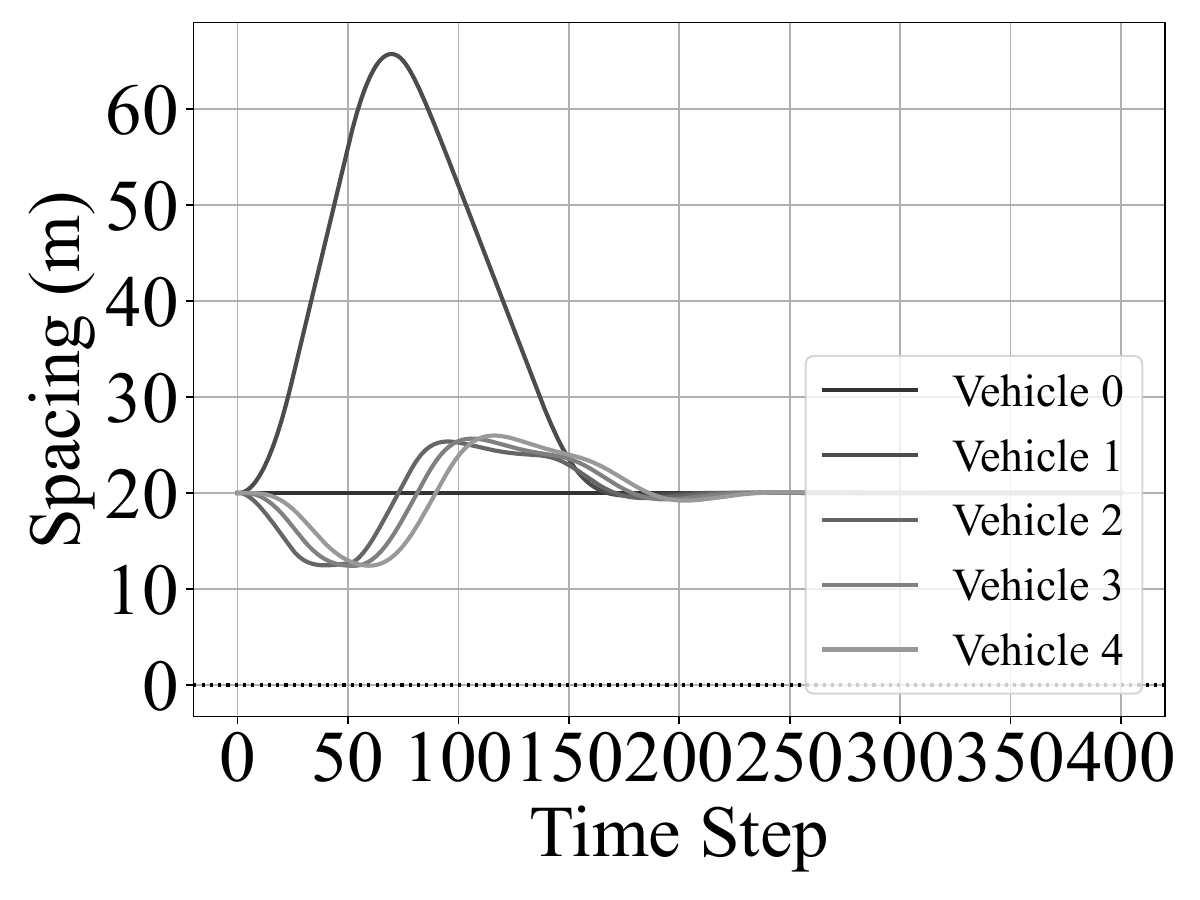}
        }
    \subcaptionbox{Velocity}{
            \includegraphics[width=5cm]{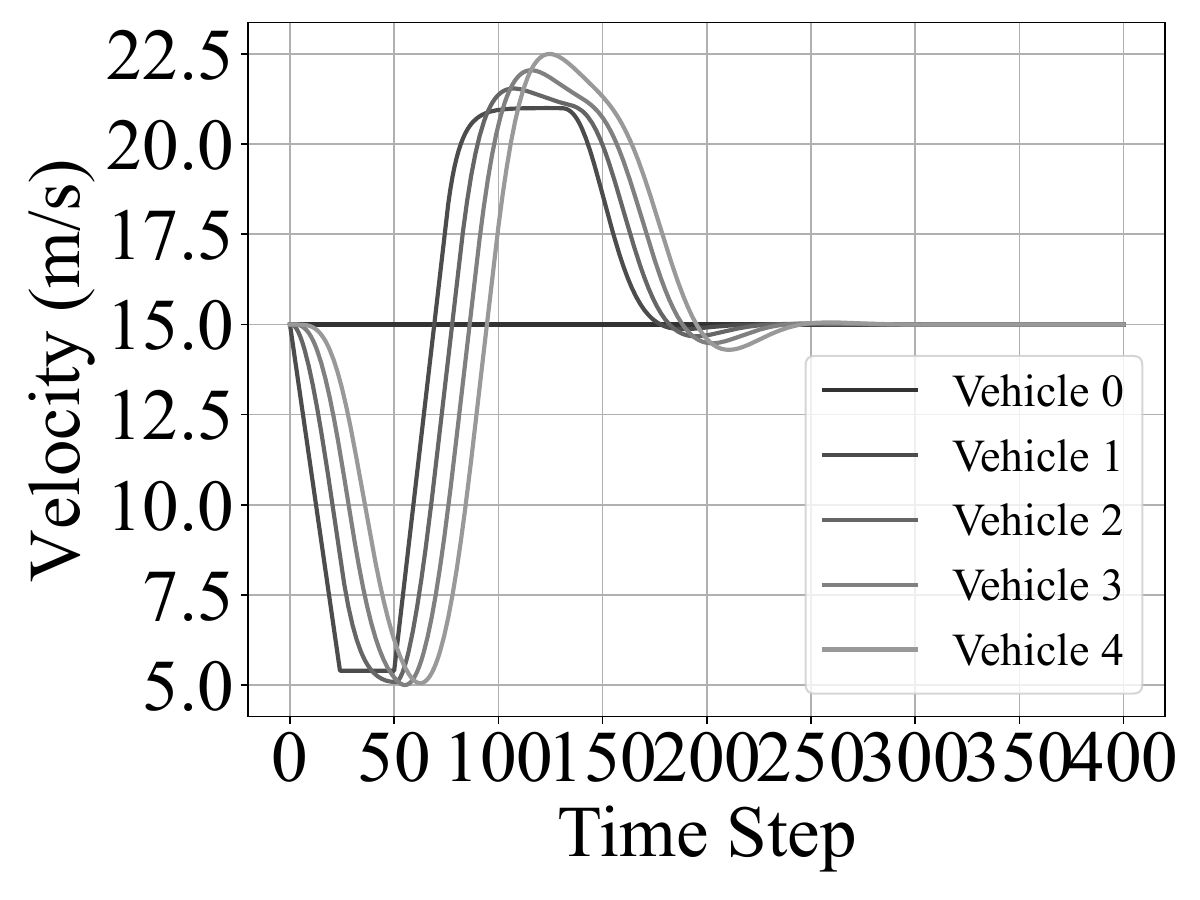}
        }
    \subcaptionbox{Acceleration}{
            \includegraphics[width=5cm]{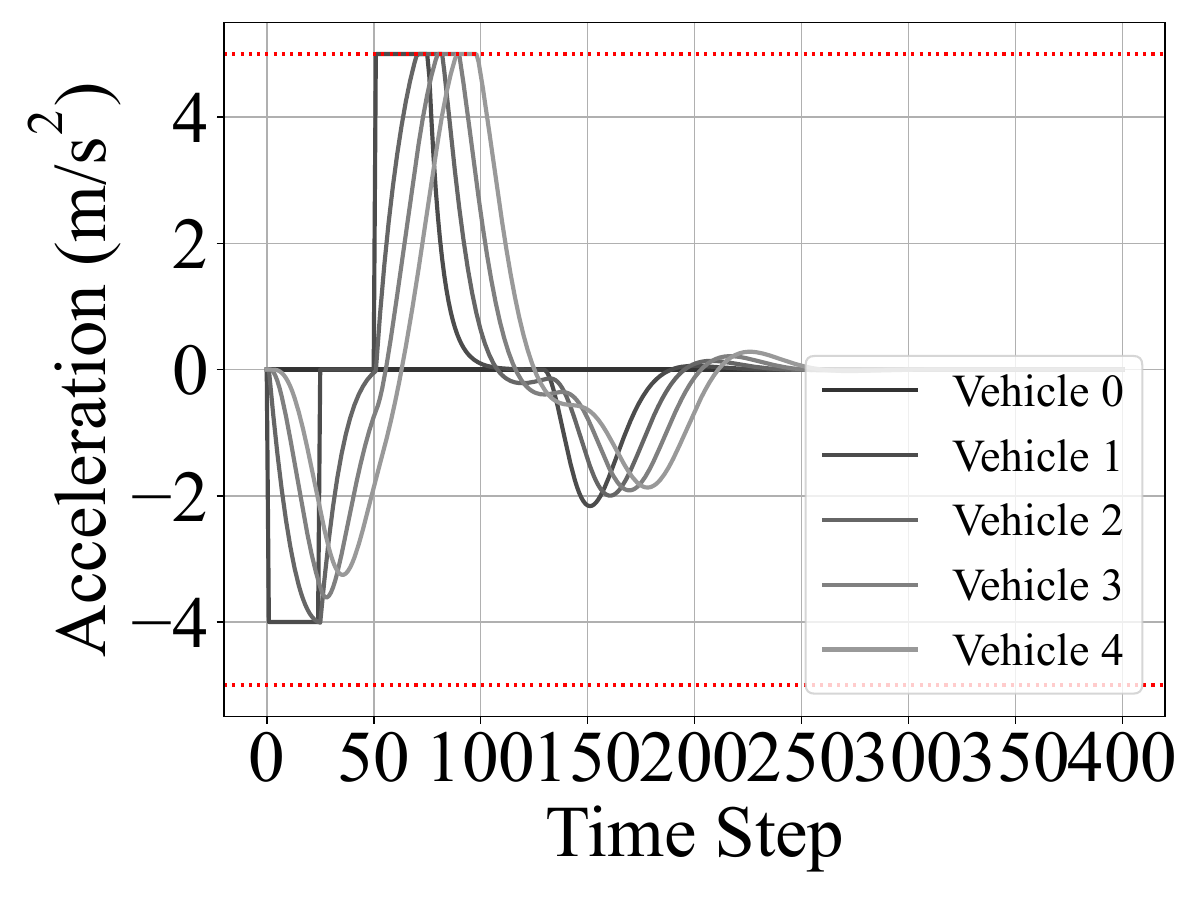}
        }
    \caption{Spacing, velocity and acceleration for vehicles in scenario 1. (a), (b), (c) are the results for safe-RL with SI. (d), (e), (f) correspond to the results for PPO without safety guarantee. The results for the pure car-following scenario are presented in (g), (h), (i).}
    \label{fig: States for scenario 1}
\end{figure*}

We simulate an emergency that occurs at $0$ s, such that a preceding HDV (with index 0 or 1) decelerates with $\rm -4m/s^2$ for $2.5$s,  maintains a low velocity for another $2.5$s, and then accelerates to equilibrium speed. Note that since the perturbation only occurs on preceding vehicles that cannot be influenced by the CAV, this scenario does not require considering HDV safety, and hence the online human driver behavior identification module does not exert any influence on the outcomes. Consequently, we only compare three models: (i) pure car following, (ii) PPO without safety guarantee, and (iii) safe-RL with SI. 
Fig.~\ref{fig: CAV safety} shows the value of the CBF candidate, and Fig.~\ref{fig: States for scenario 1} shows the spacing, velocity, and acceleration at each time step in the test simulation. For pure car following and safe-RL with SI, the resulting spacing between the CAV and the preceding HDV and the value of the CBF candidate are both greater than $0$, which indicates that the collision does not occur. However, safety is not guaranteed for PPO without safety guarantee, since the CAV under the algorithm collides with the preceding HDV. 
Furthermore, as illustrated in Fig.~\ref{fig: States for scenario 1} (c), the resulting control input from safe-RL with SI remains within the actuator limitation bound (depicted by the red dotted line), which is ensured by the adherence to feasibility constraint in Eq. \eqref{eq:feasibility constraint}.

\subsubsection{Case Study for Scenario 2}
\begin{figure}[ht]
    \centering
    \includegraphics[width=7cm]{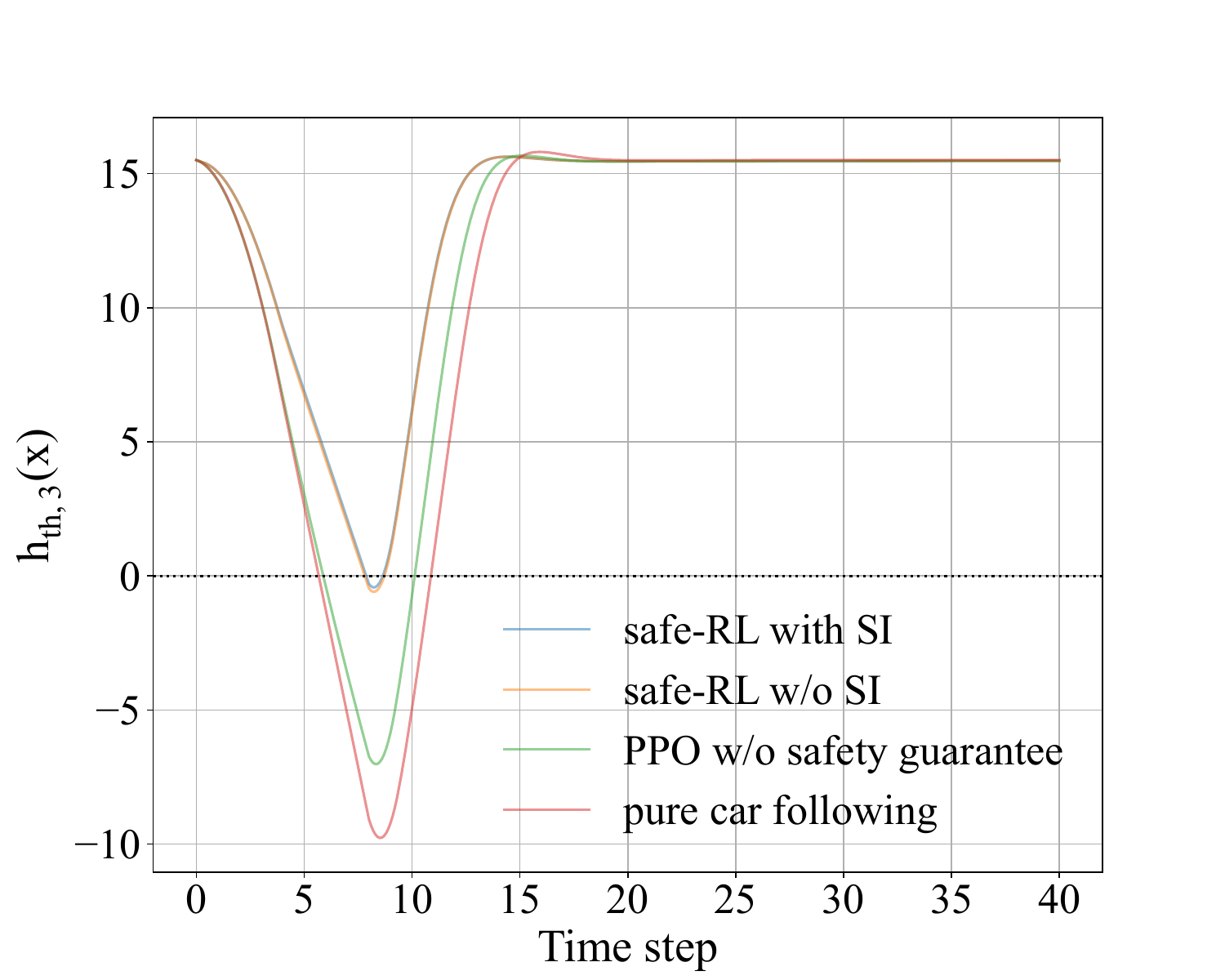}
    \caption{Value of the CBF candidate for vehicle $3$ when it accelerates emergently.}
    \label{fig: HDV safety 1}
\end{figure}
\begin{figure}[ht]
    \centering
    \includegraphics[width=7cm]{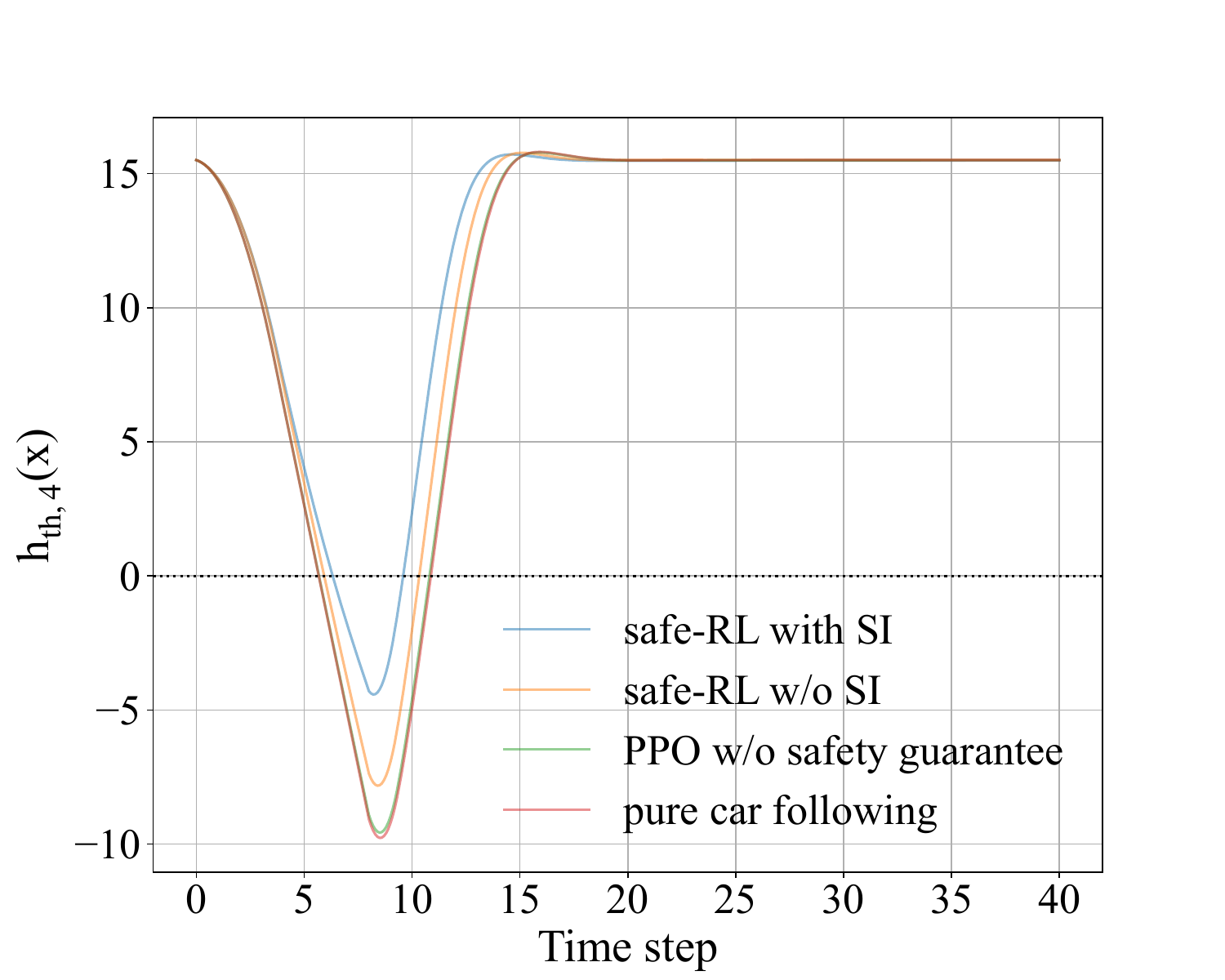}
    \caption{Value of the CBF candidate for vehicle $4$ when it accelerates emergently.}
    \label{fig: HDV safety 2}
\end{figure}
\begin{figure*}[ht]
    \centering
    \subcaptionbox{Spacing}{
            \includegraphics[width=5cm]{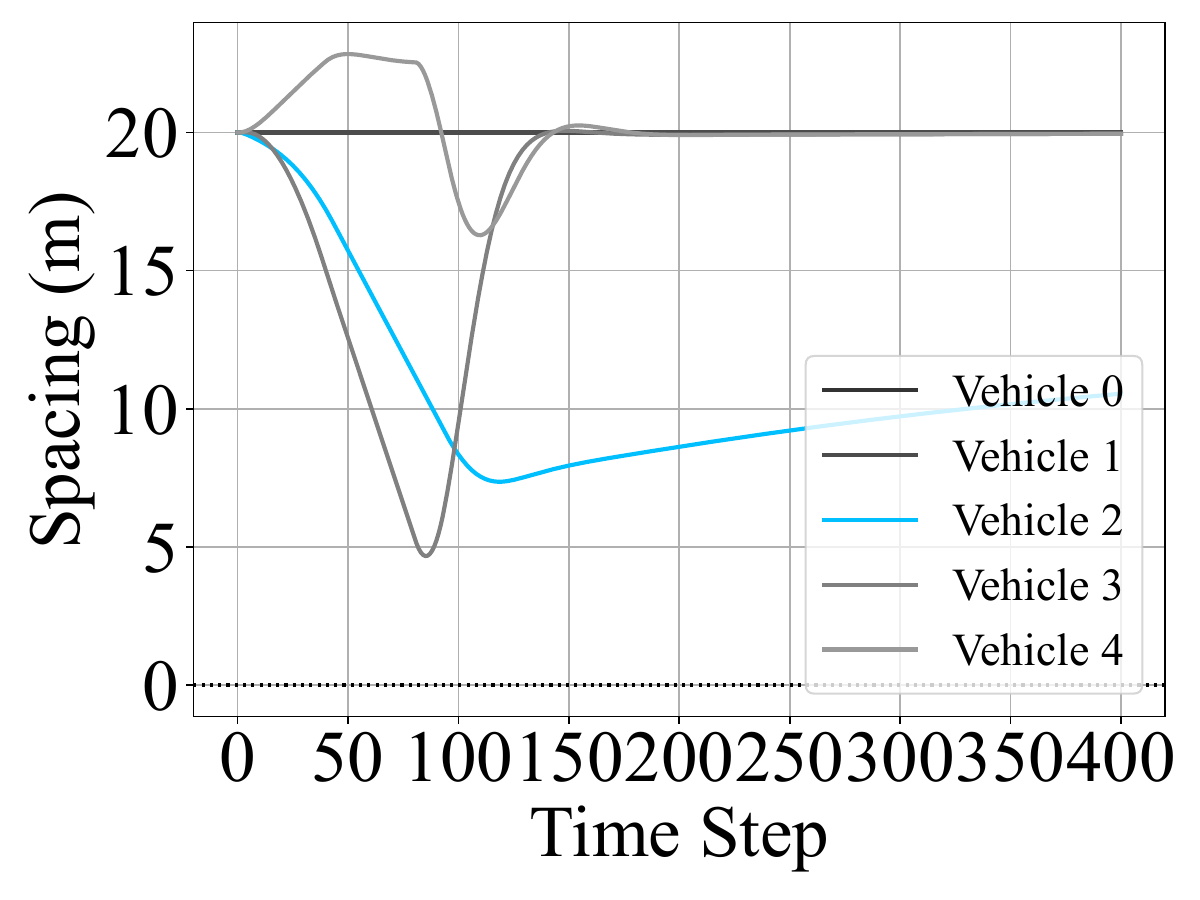}
        }
    \subcaptionbox{Velocity}{
            \includegraphics[width=5cm]{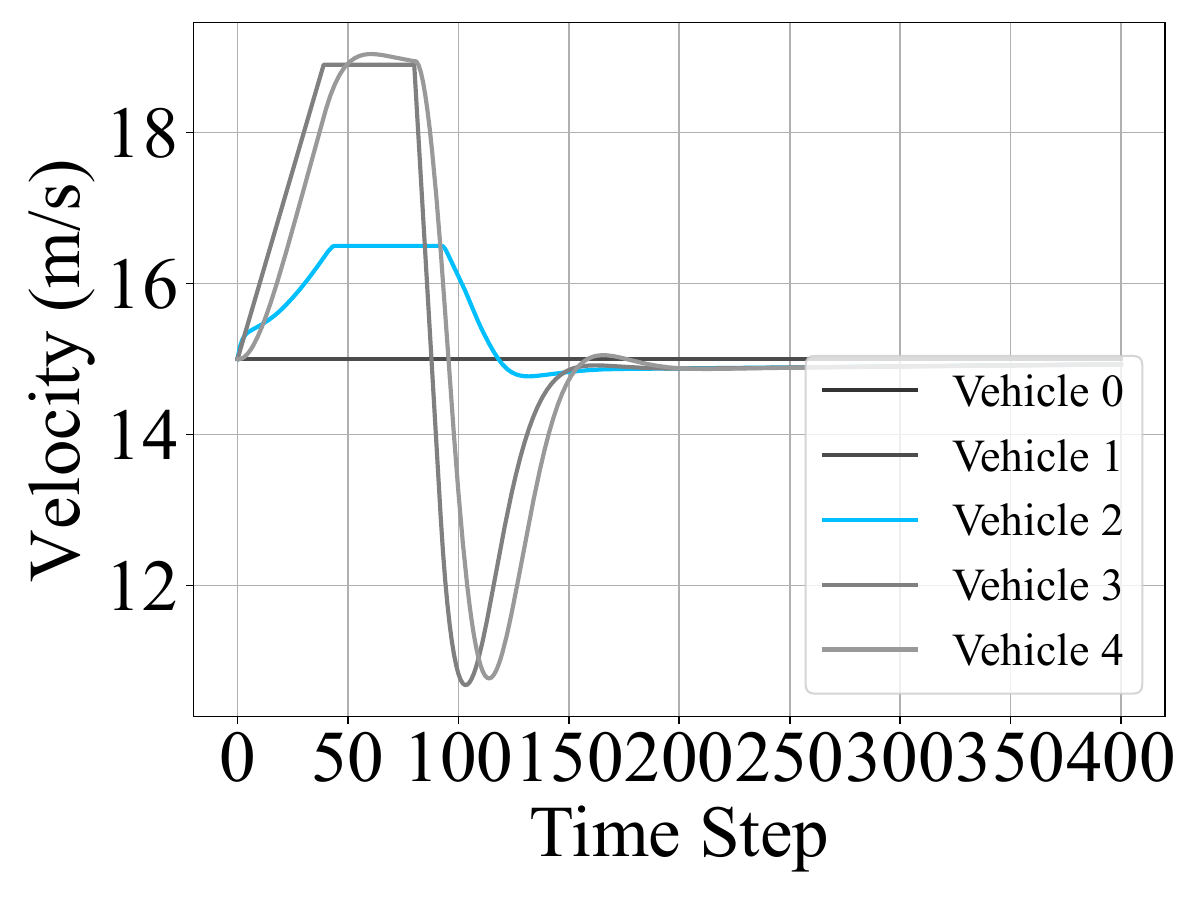}
        }
    \subcaptionbox{Acceleration}{
            \includegraphics[width=5cm]{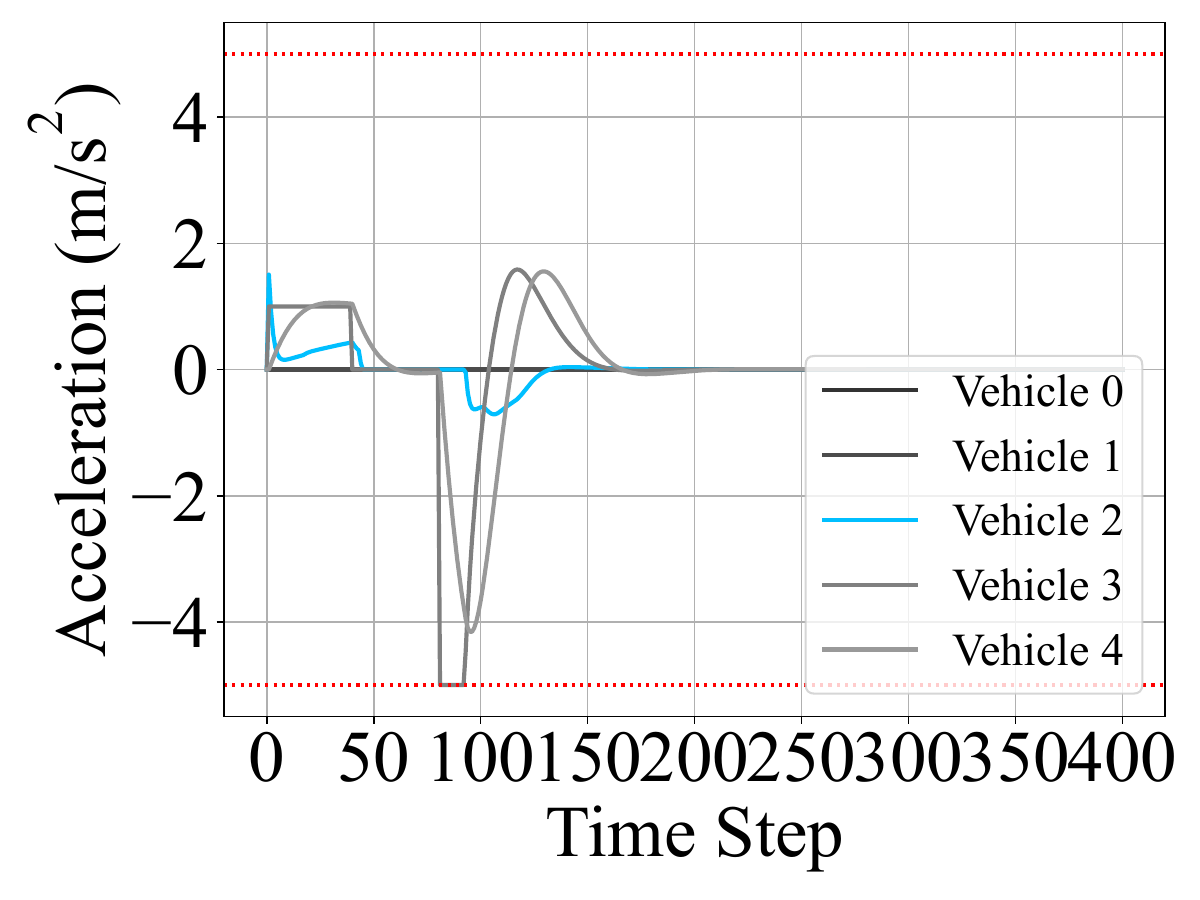}
        }
    \subcaptionbox{Spacing}{
            \includegraphics[width=5cm]{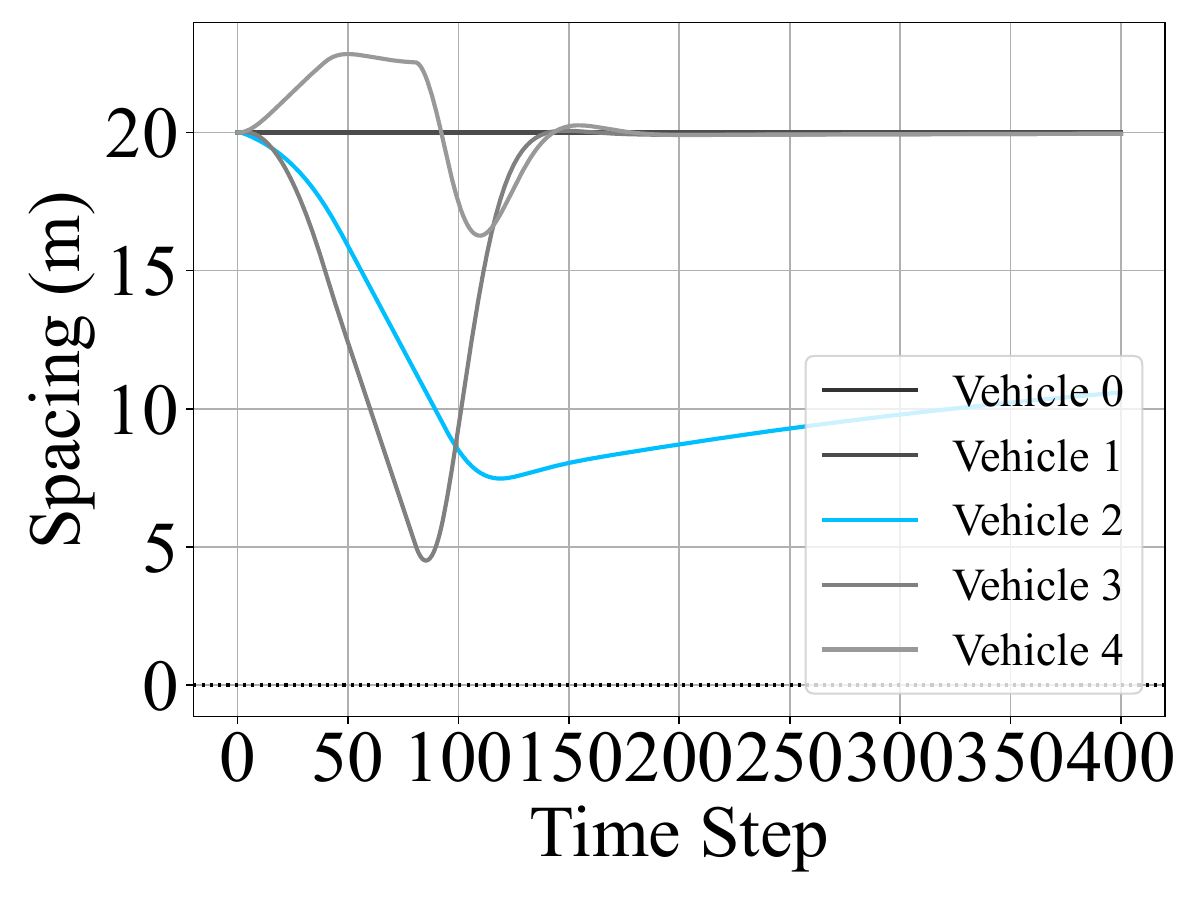}
        }
    \subcaptionbox{Velocity}{
            \includegraphics[width=5cm]{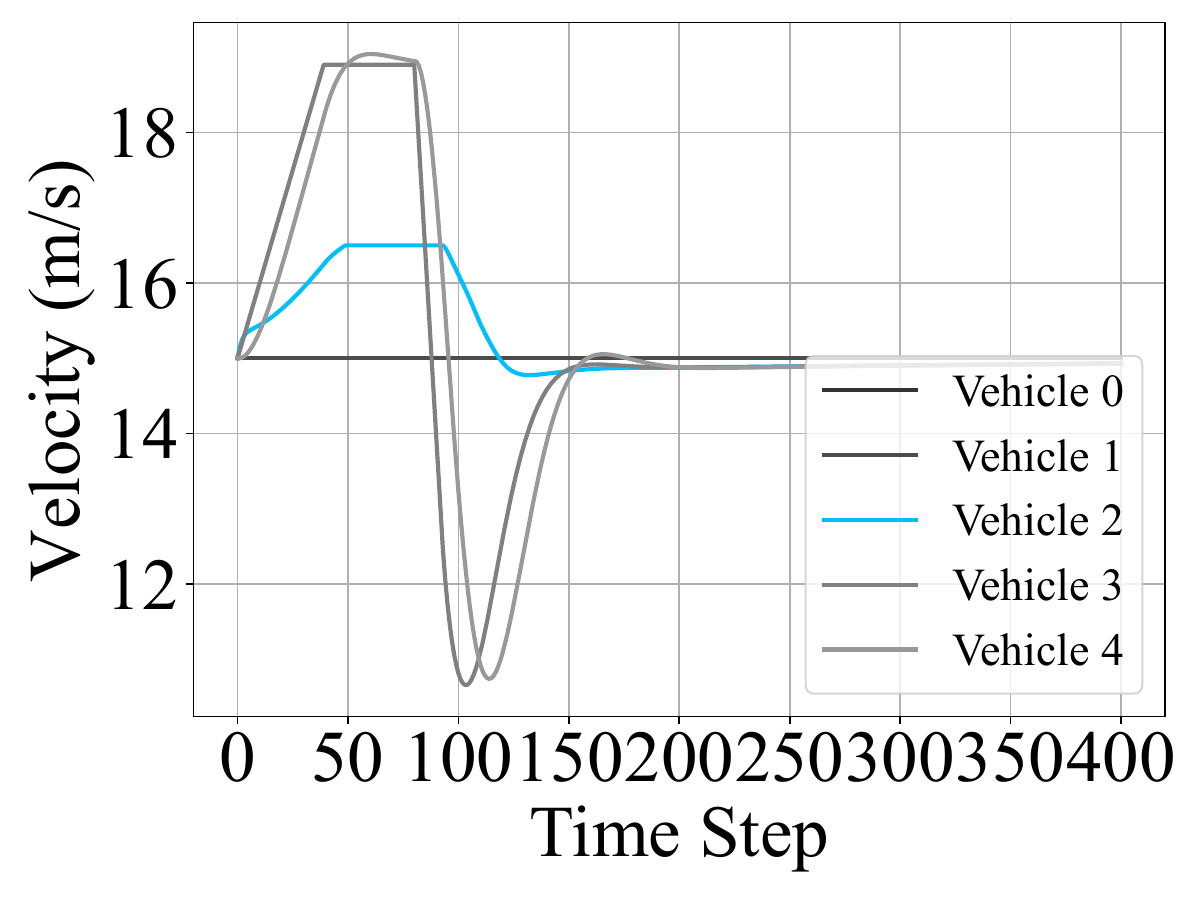}
        }
    \subcaptionbox{Acceleration}{
            \includegraphics[width=5cm]{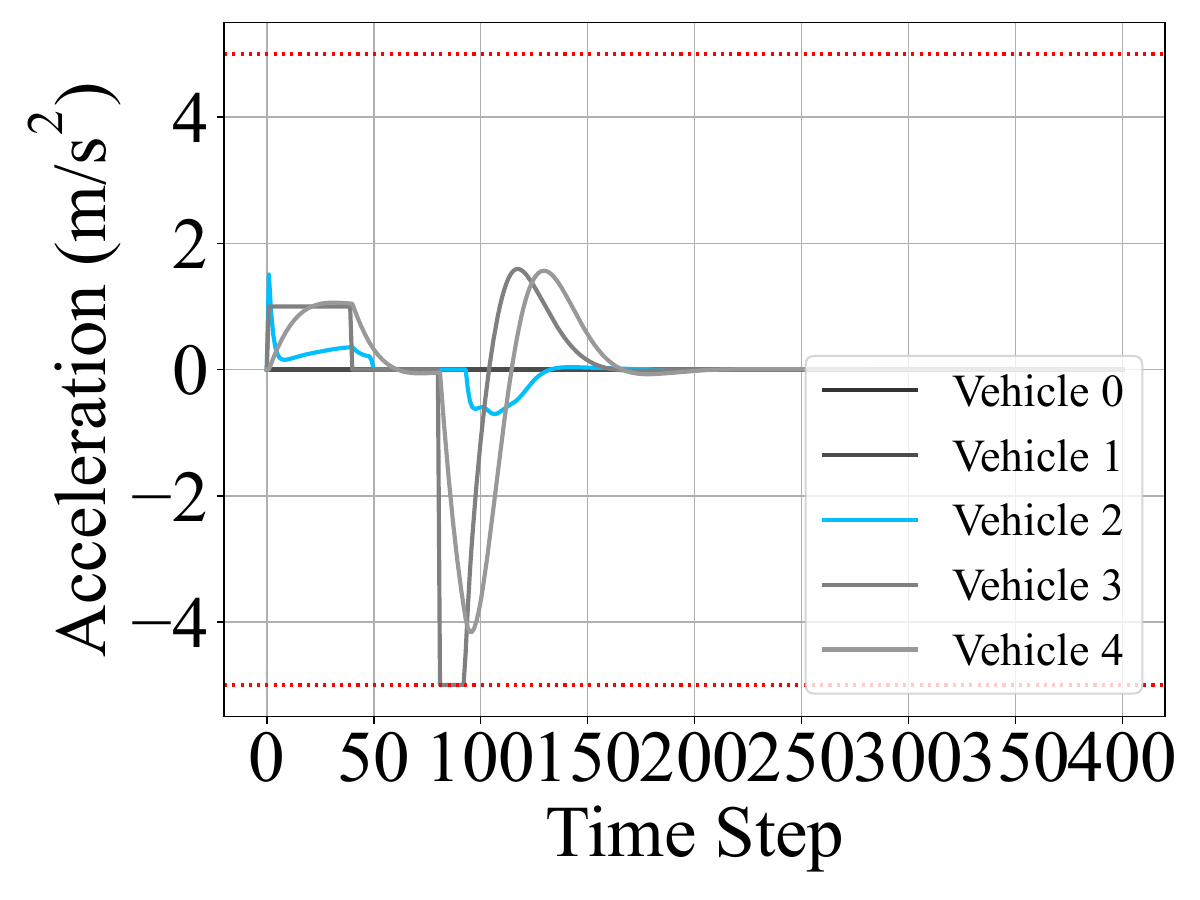}
        }
    \subcaptionbox{Spacing}{
            \includegraphics[width=5cm]{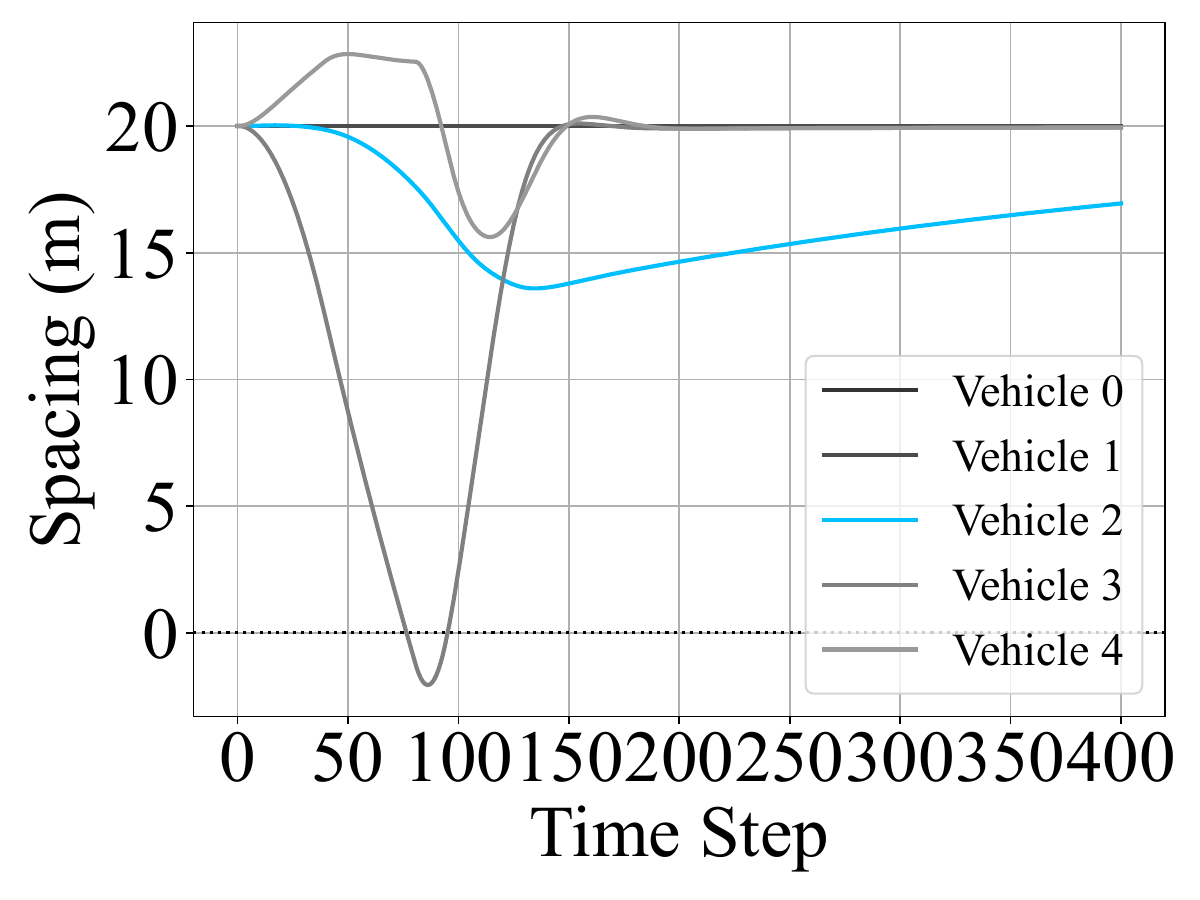}
        }
    \subcaptionbox{Velocity}{
            \includegraphics[width=5cm]{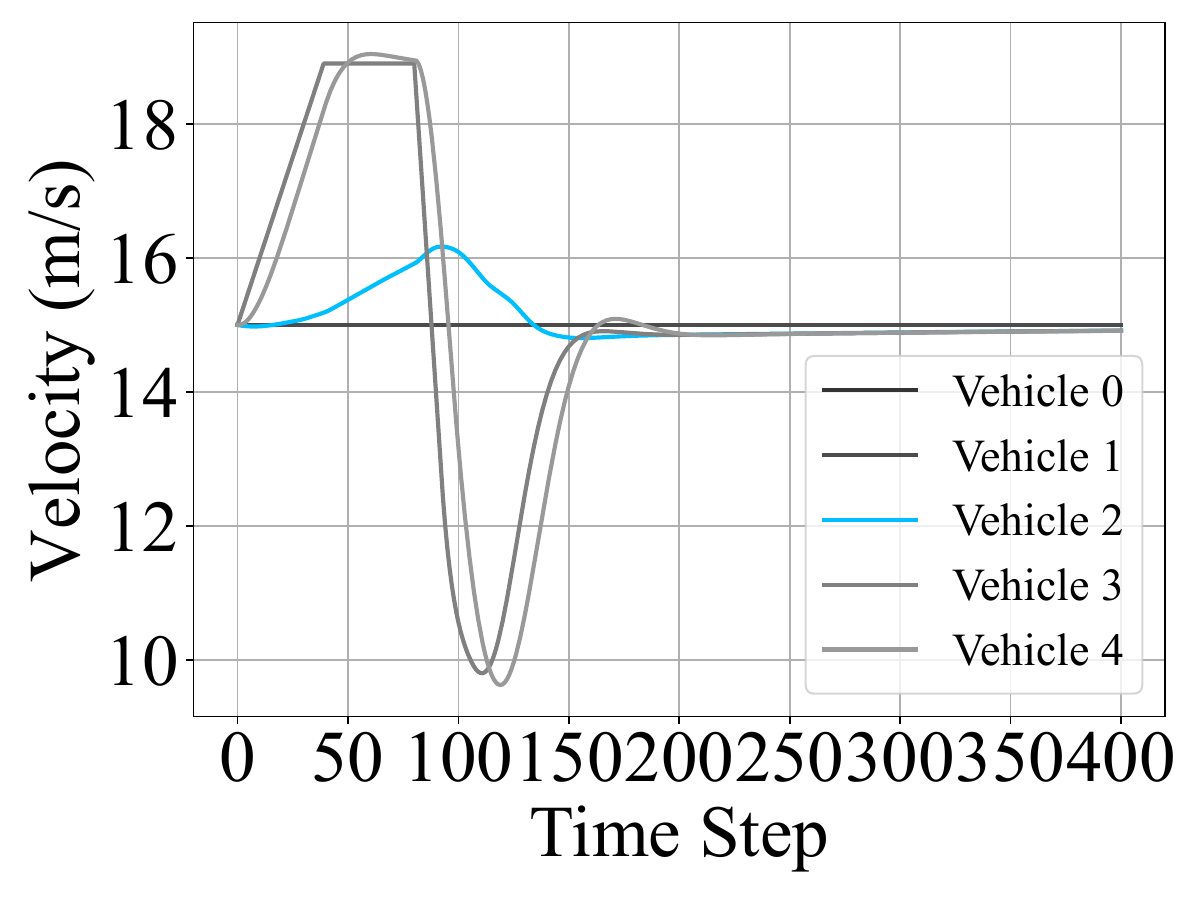}
        }
    \subcaptionbox{Acceleration}{
            \includegraphics[width=5cm]{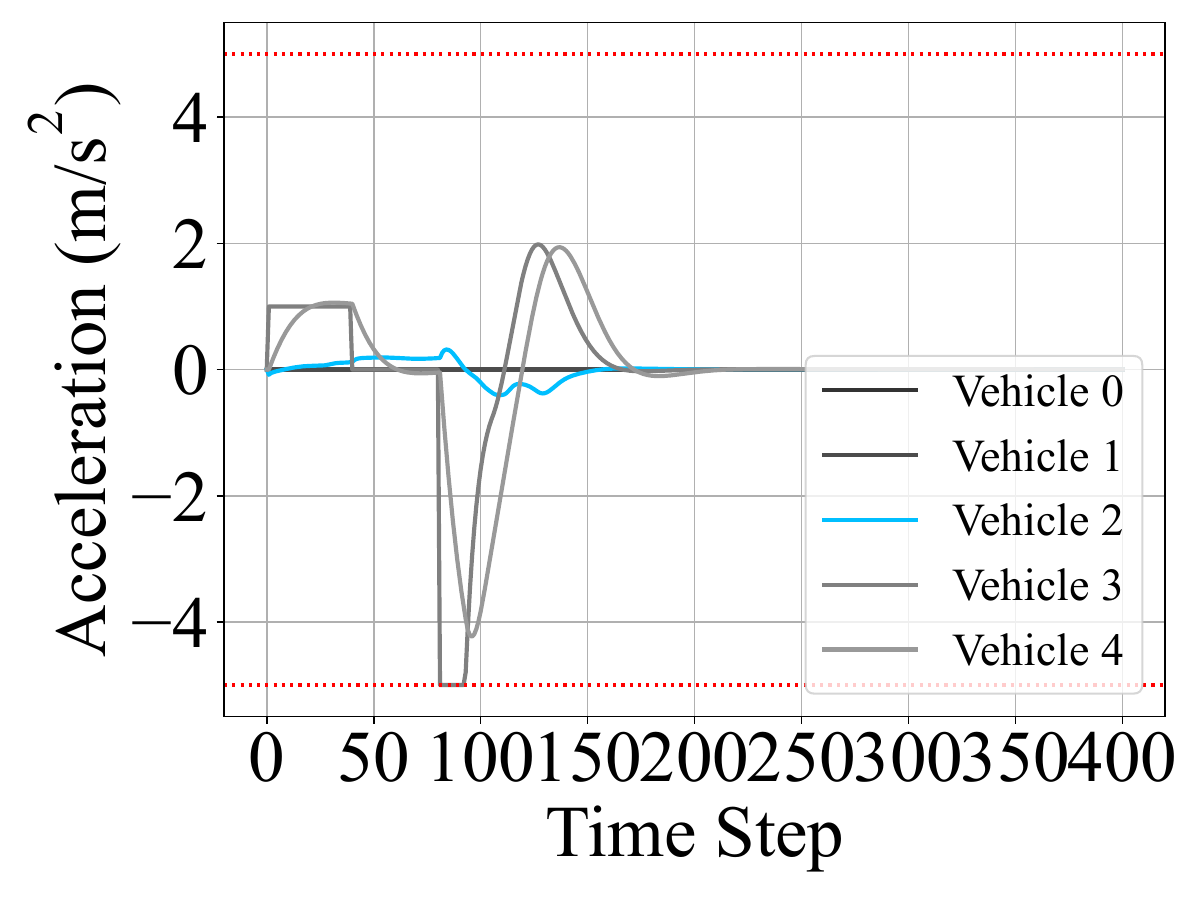}
        }
    \subcaptionbox{Spacing}{
            \includegraphics[width=5cm]{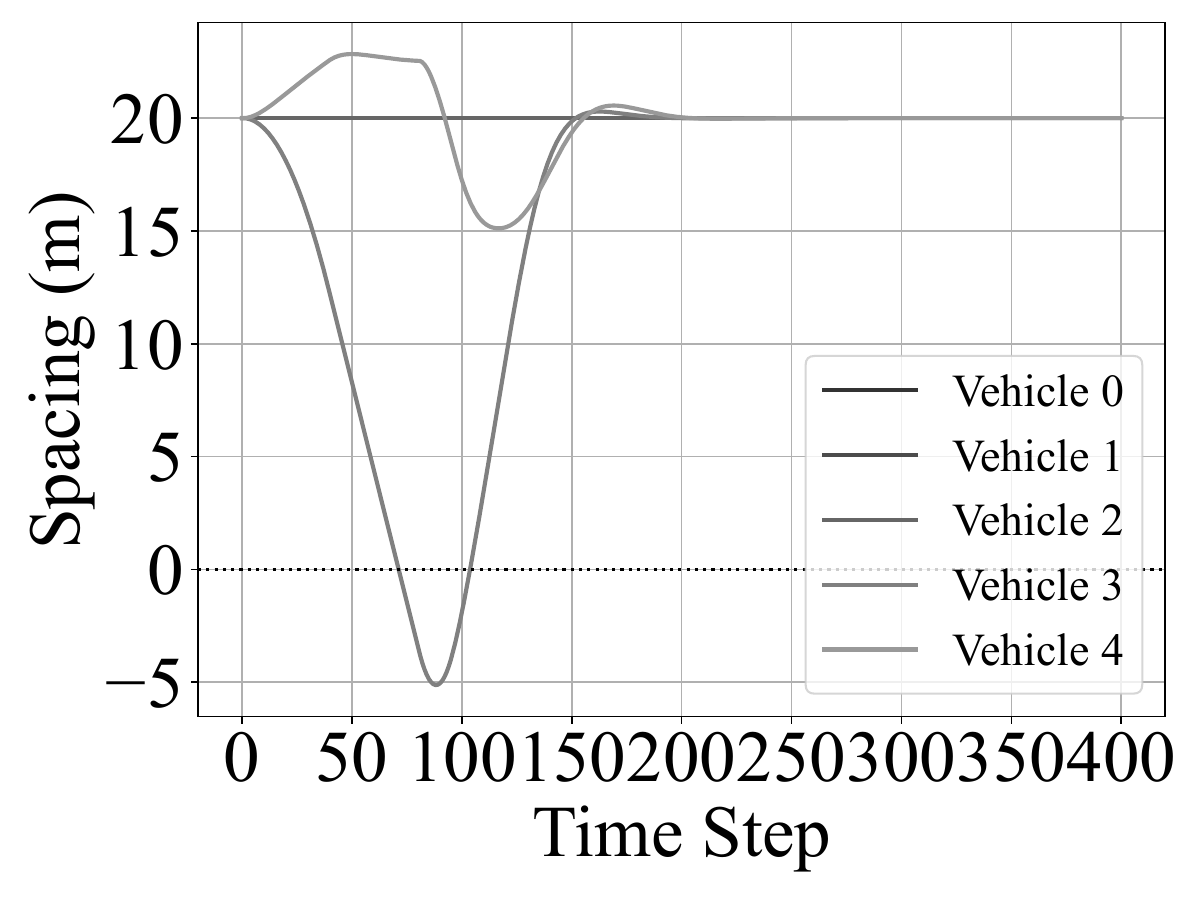}
        }
    \subcaptionbox{Velocity}{
            \includegraphics[width=5cm]{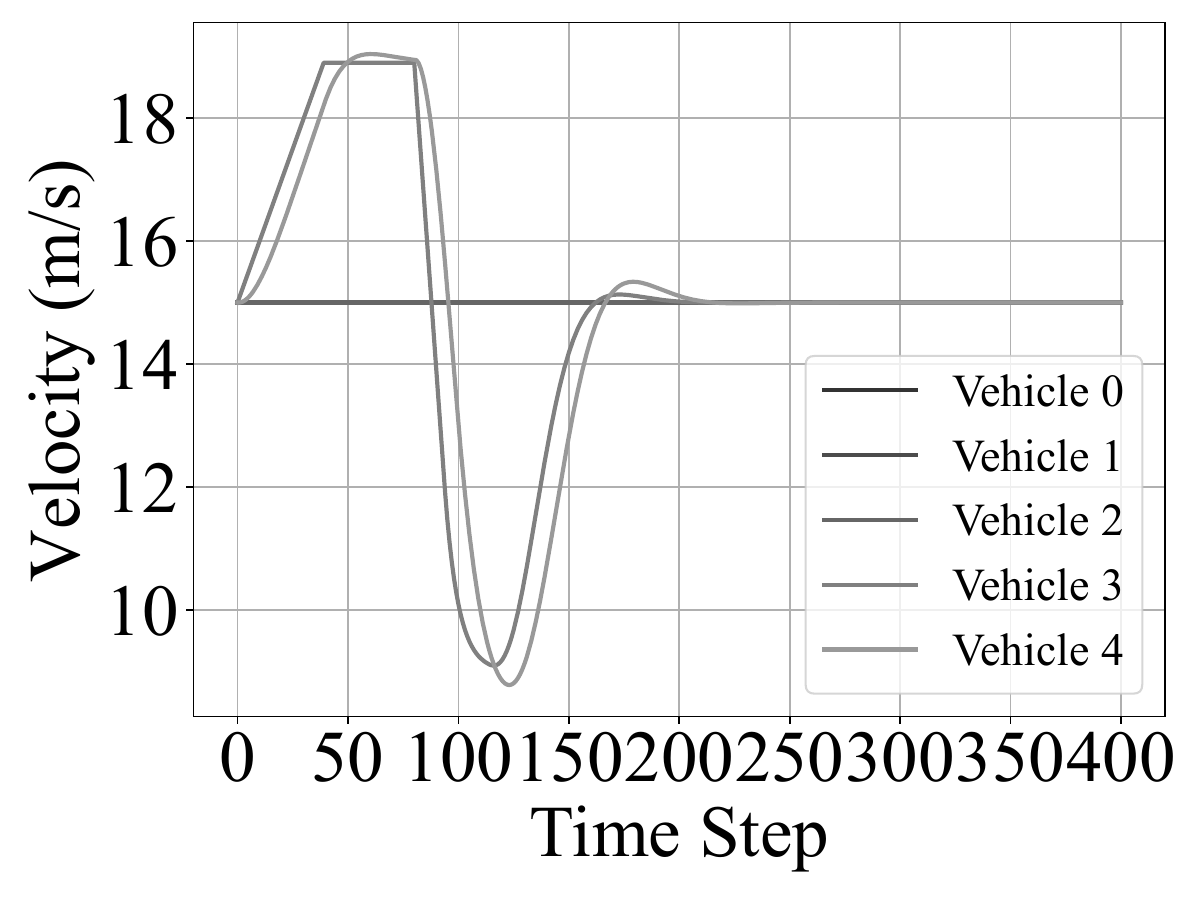}
        }
    \subcaptionbox{Acceleration}{
            \includegraphics[width=5cm]{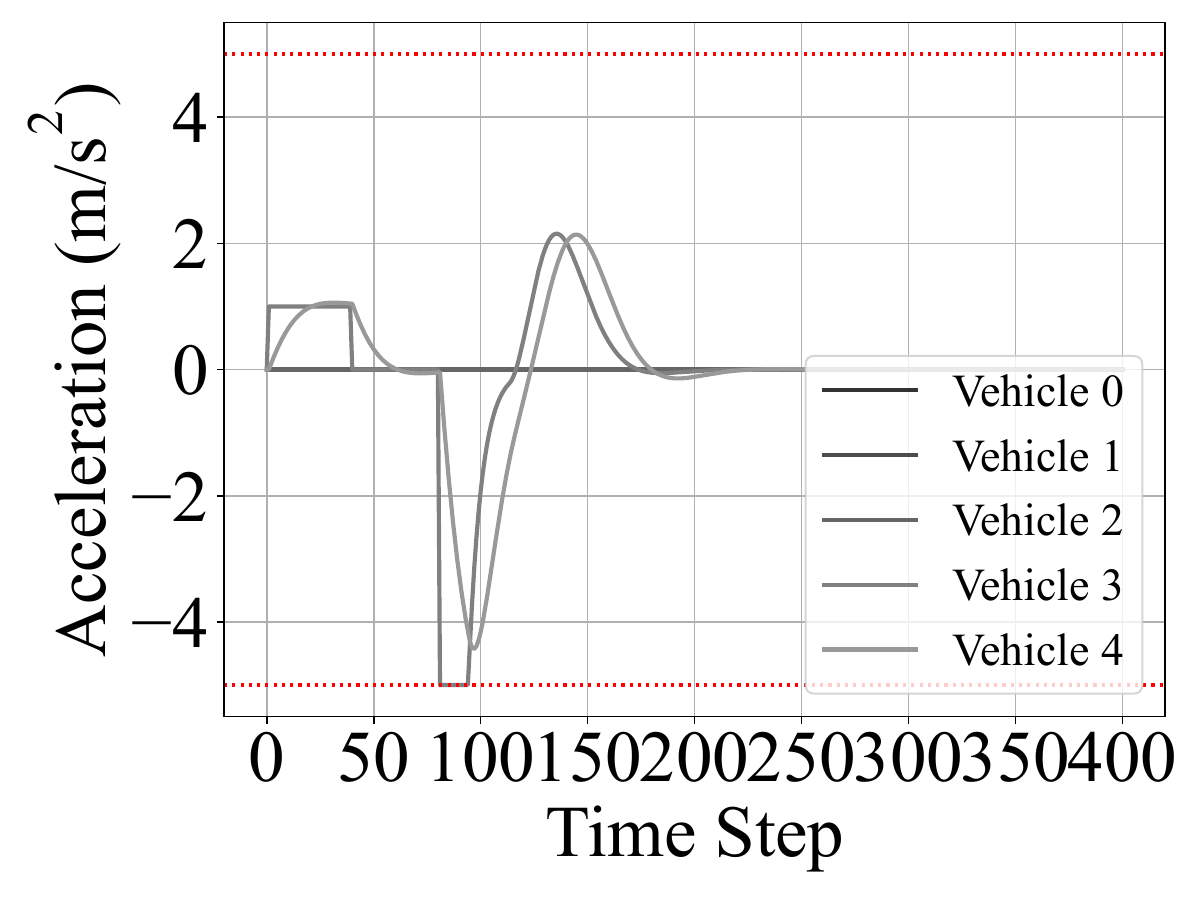}
        }
    \caption{Spacing, velocity and acceleration for vehicles in scenario 2 when vehicle 4 accelerates. (a), (b), (c) are the results for safe-RL with SI. (d), (e), (f) show the results for safe-RL without SI. (g), (h), (i) correspond to the results for PPO without safety guarantee. The results for the pure car-following scenario are presented in (j), (k), (l).}
    \label{fig: States for scenario 2-1}
\end{figure*}
\begin{figure*}[ht]
    \centering
    \subcaptionbox{Spacing}{
            \includegraphics[width=5cm]{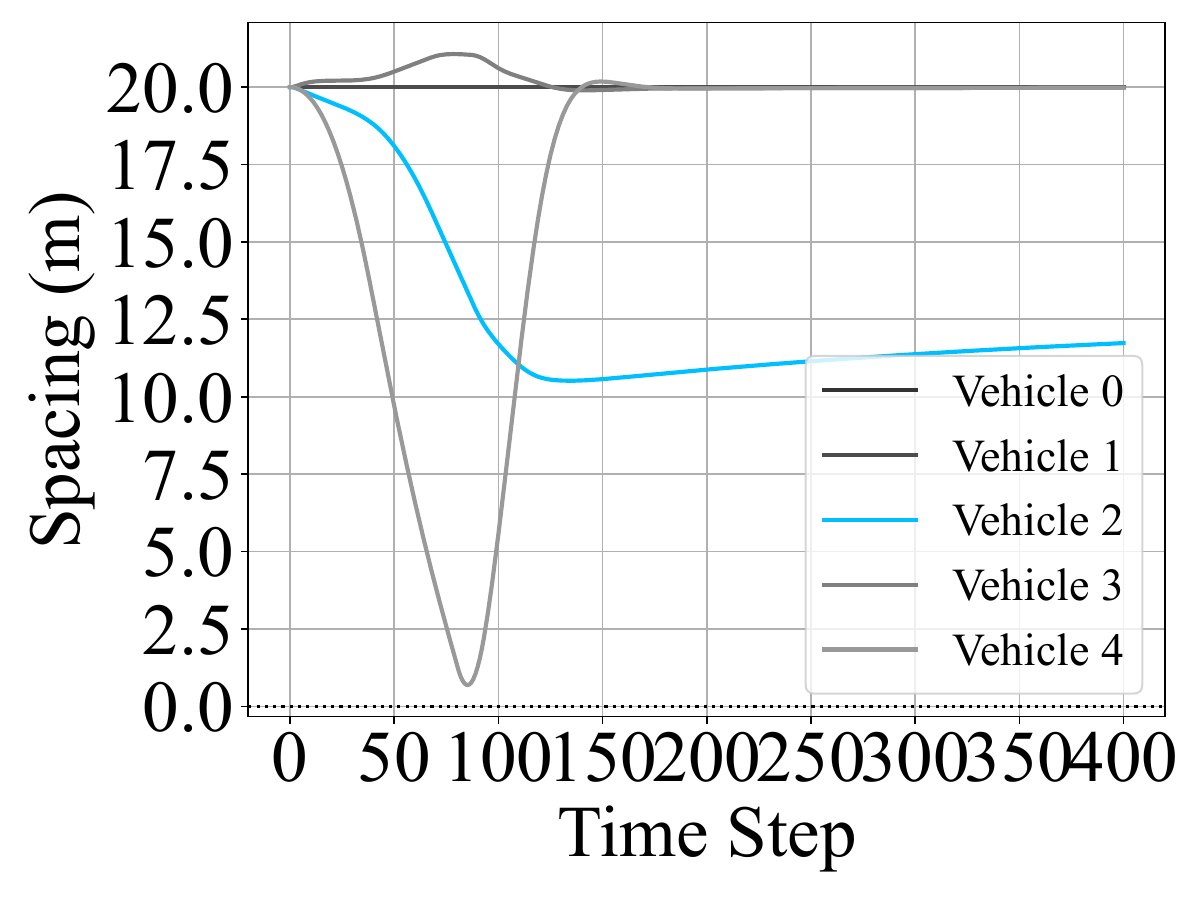}
        }
    \subcaptionbox{Velocity}{
            \includegraphics[width=5cm]{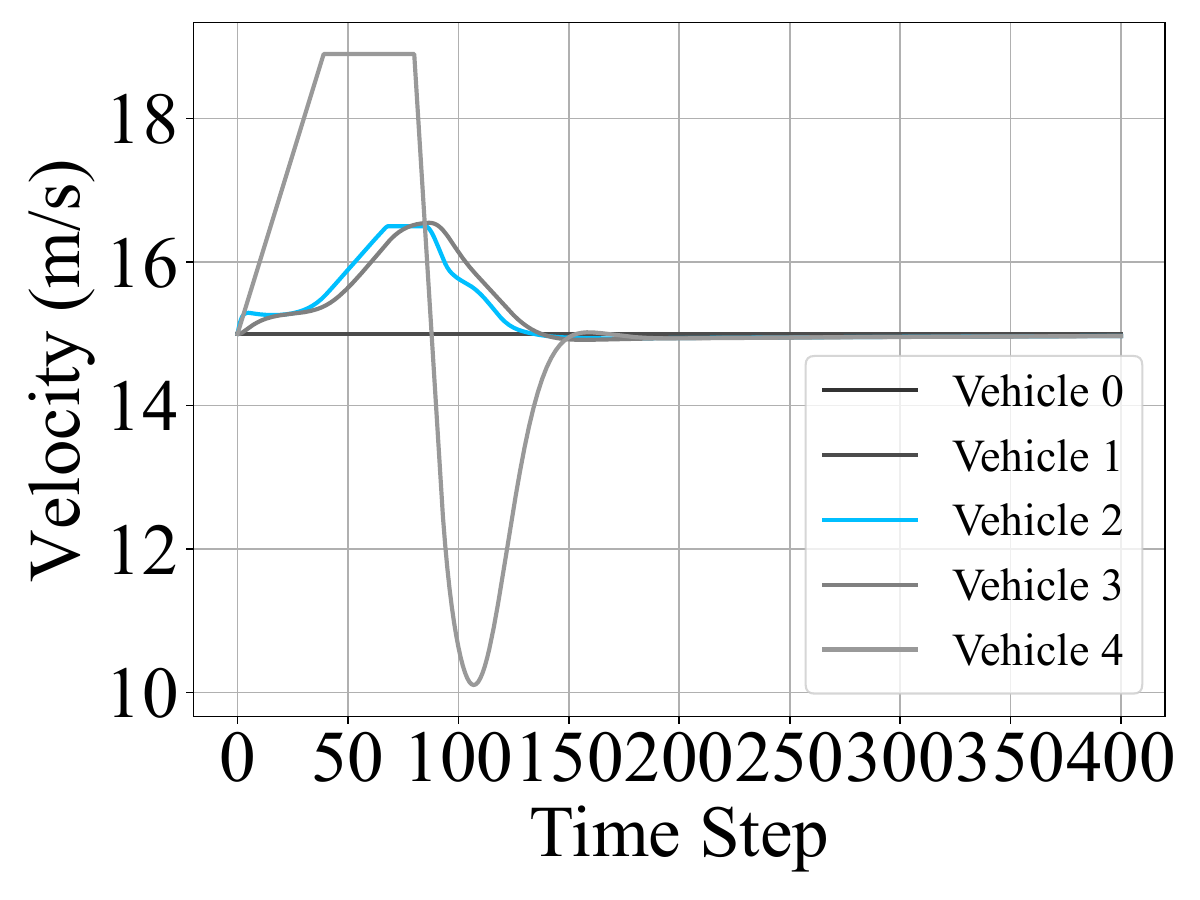}
        }
    \subcaptionbox{Acceleration}{
            \includegraphics[width=5cm]{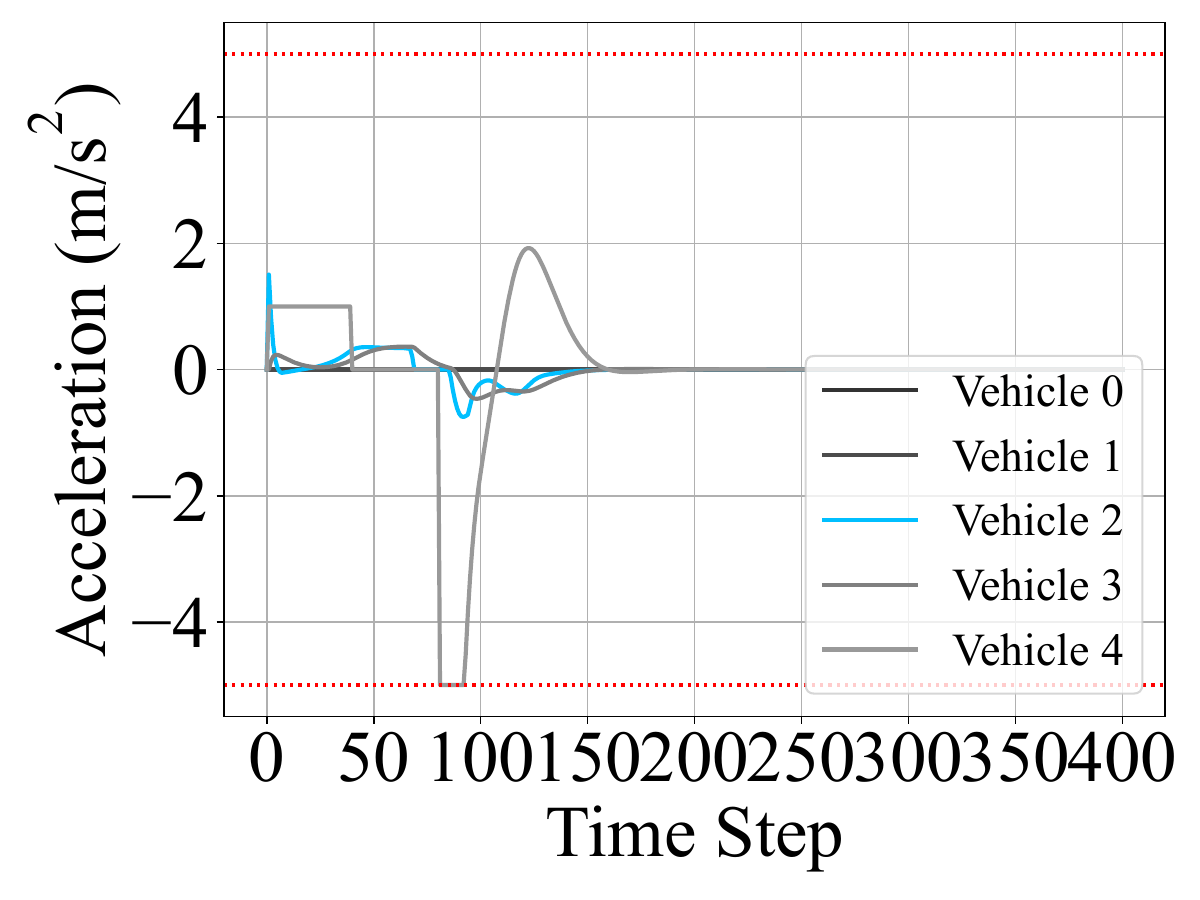}
        }
    \subcaptionbox{Spacing}{
            \includegraphics[width=5cm]{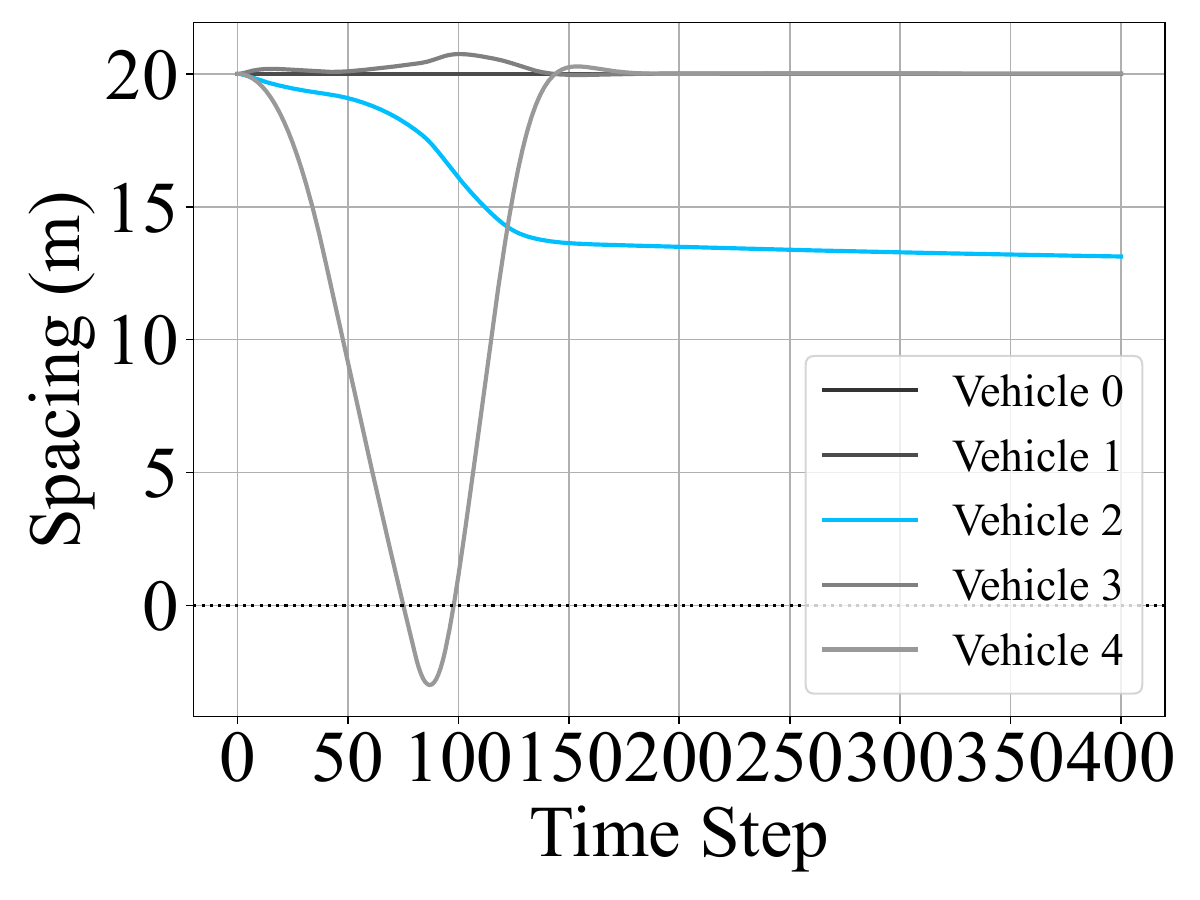}
        }
    \subcaptionbox{Velocity}{
            \includegraphics[width=5cm]{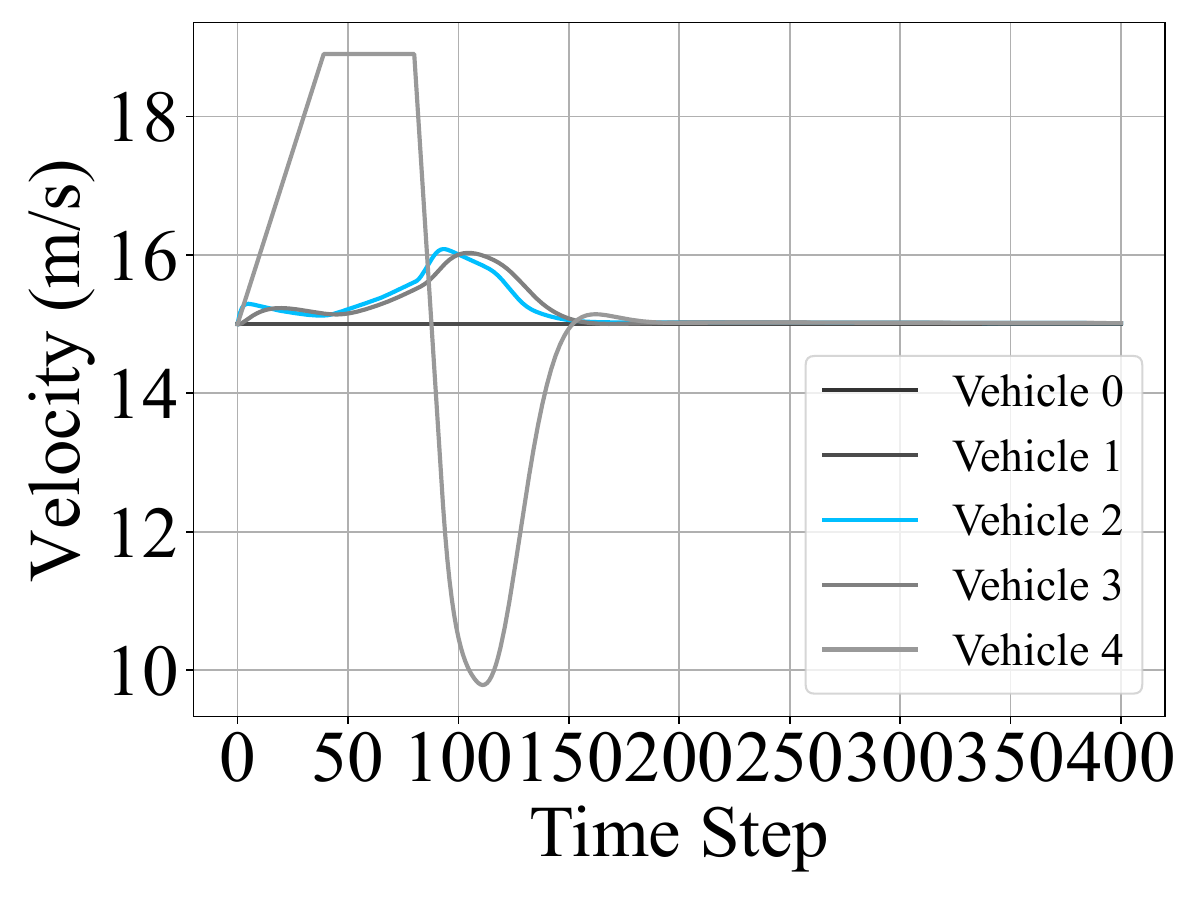}
        }
    \subcaptionbox{Acceleration}{
            \includegraphics[width=5cm]{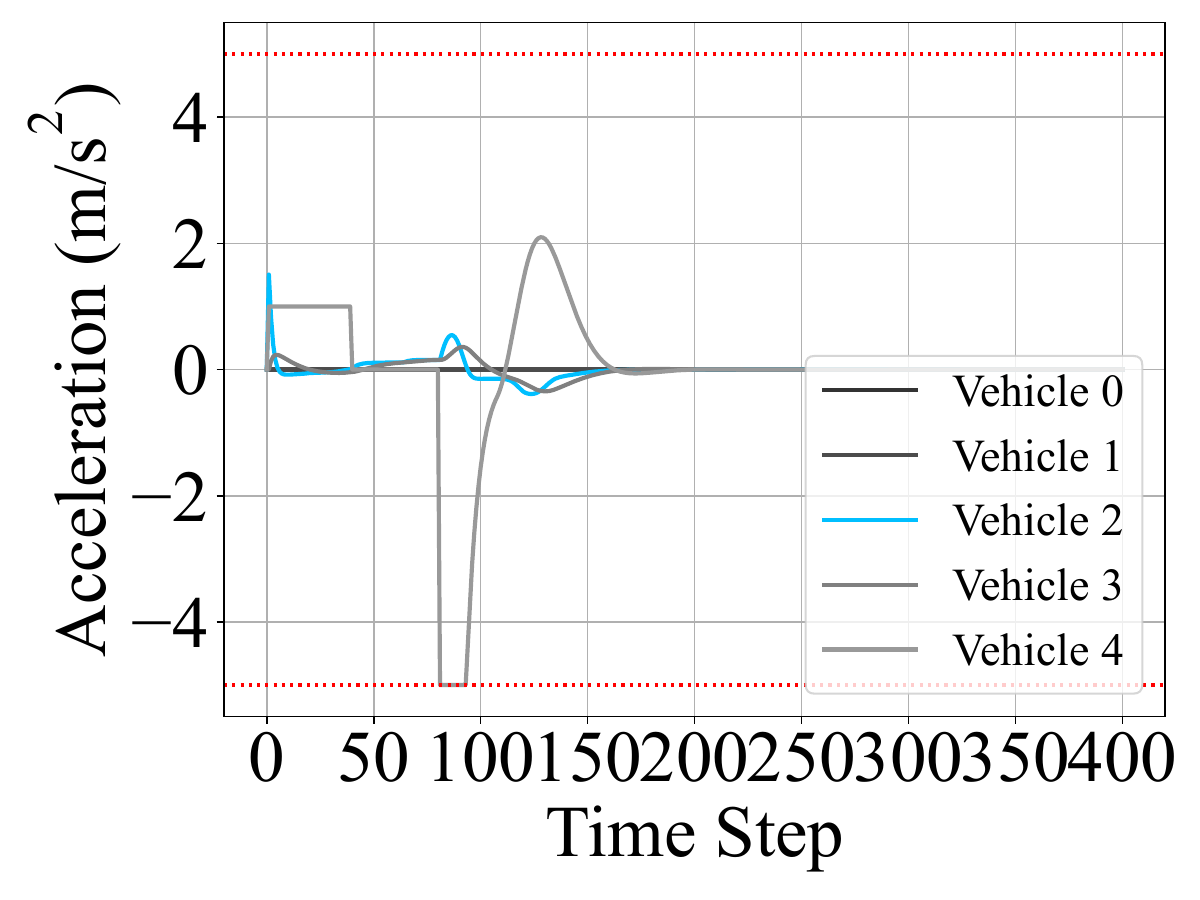}
        }
    \subcaptionbox{Spacing}{
            \includegraphics[width=5cm]{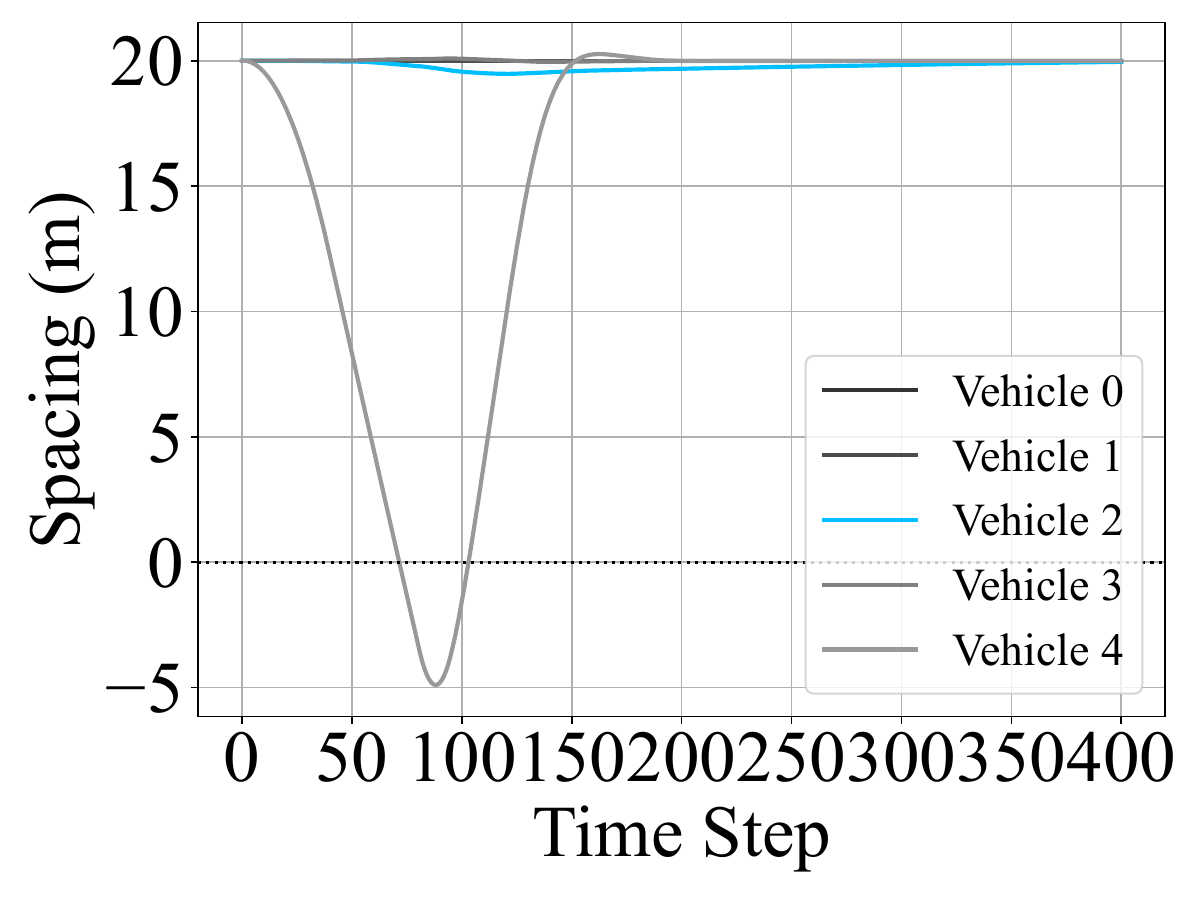}
        }
    \subcaptionbox{Velocity}{
            \includegraphics[width=5cm]{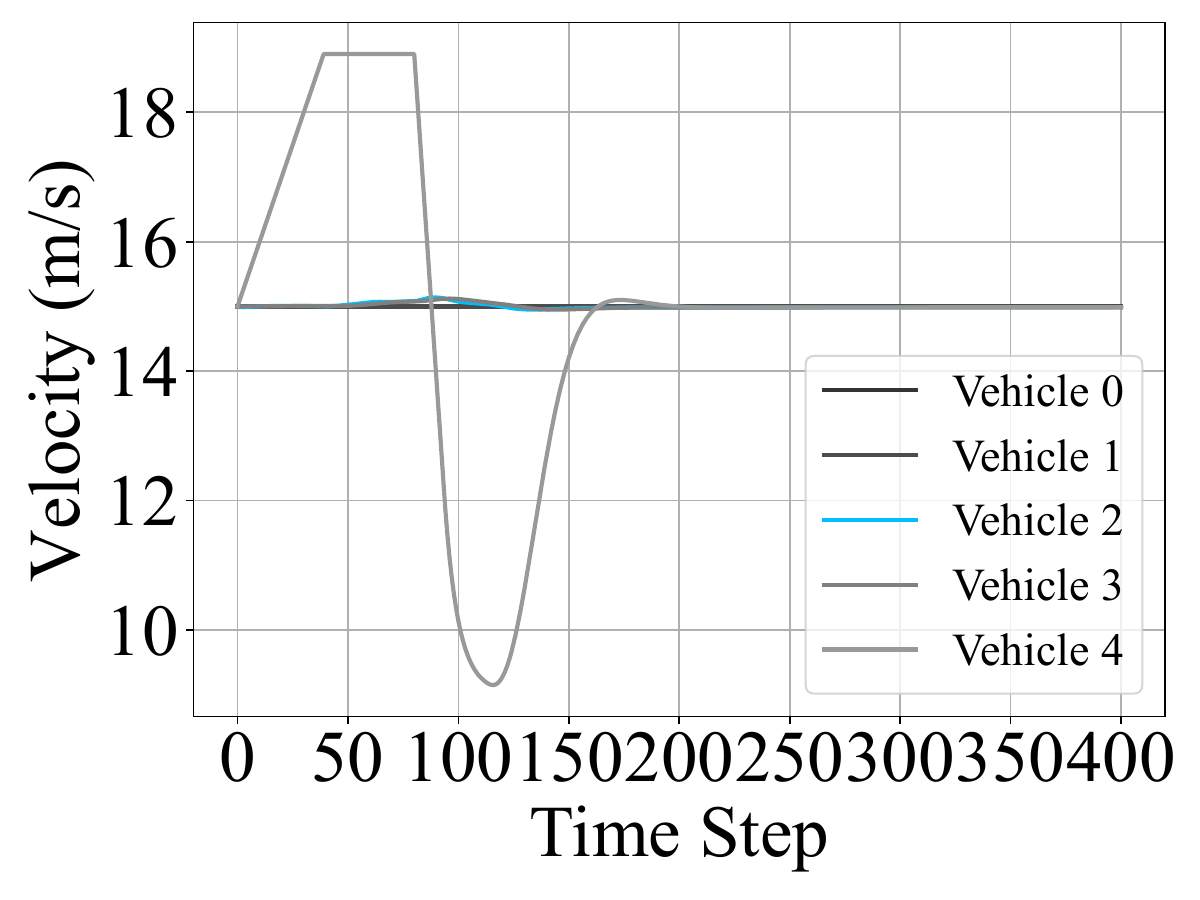}
        }
    \subcaptionbox{Acceleration}{
            \includegraphics[width=5cm]{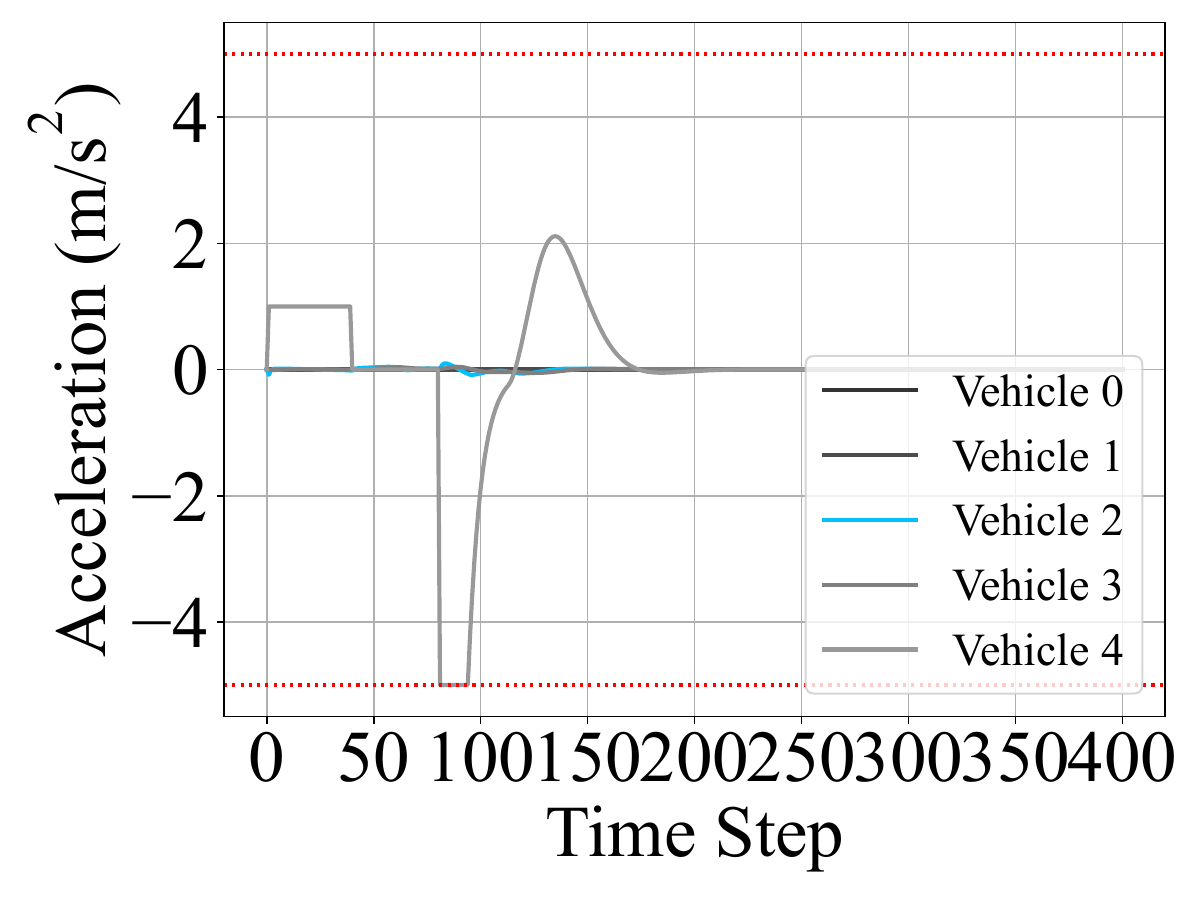}
        }
    \subcaptionbox{Spacing}{
            \includegraphics[width=5cm]{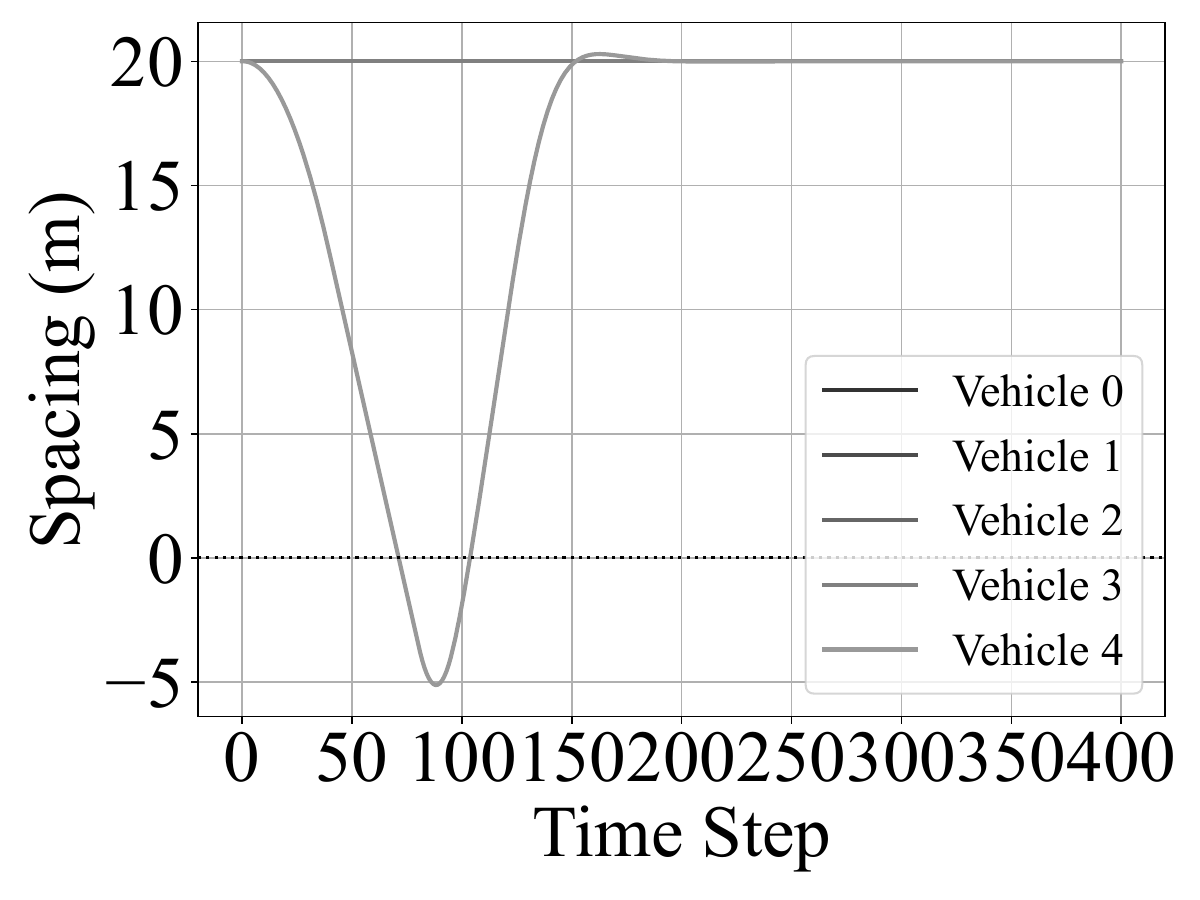}
        }
    \subcaptionbox{Velocity}{
            \includegraphics[width=5cm]{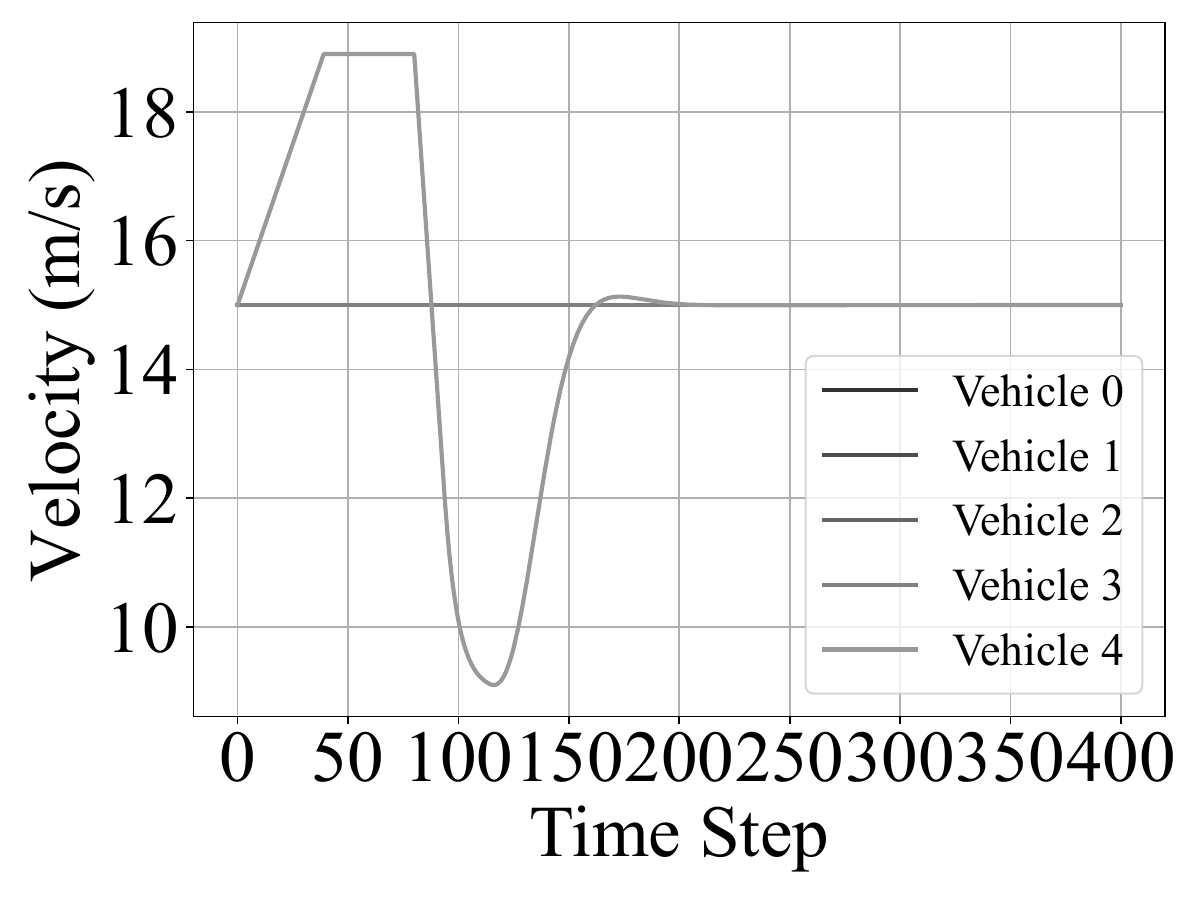}
        }
    \subcaptionbox{Acceleration}{
            \includegraphics[width=5cm]{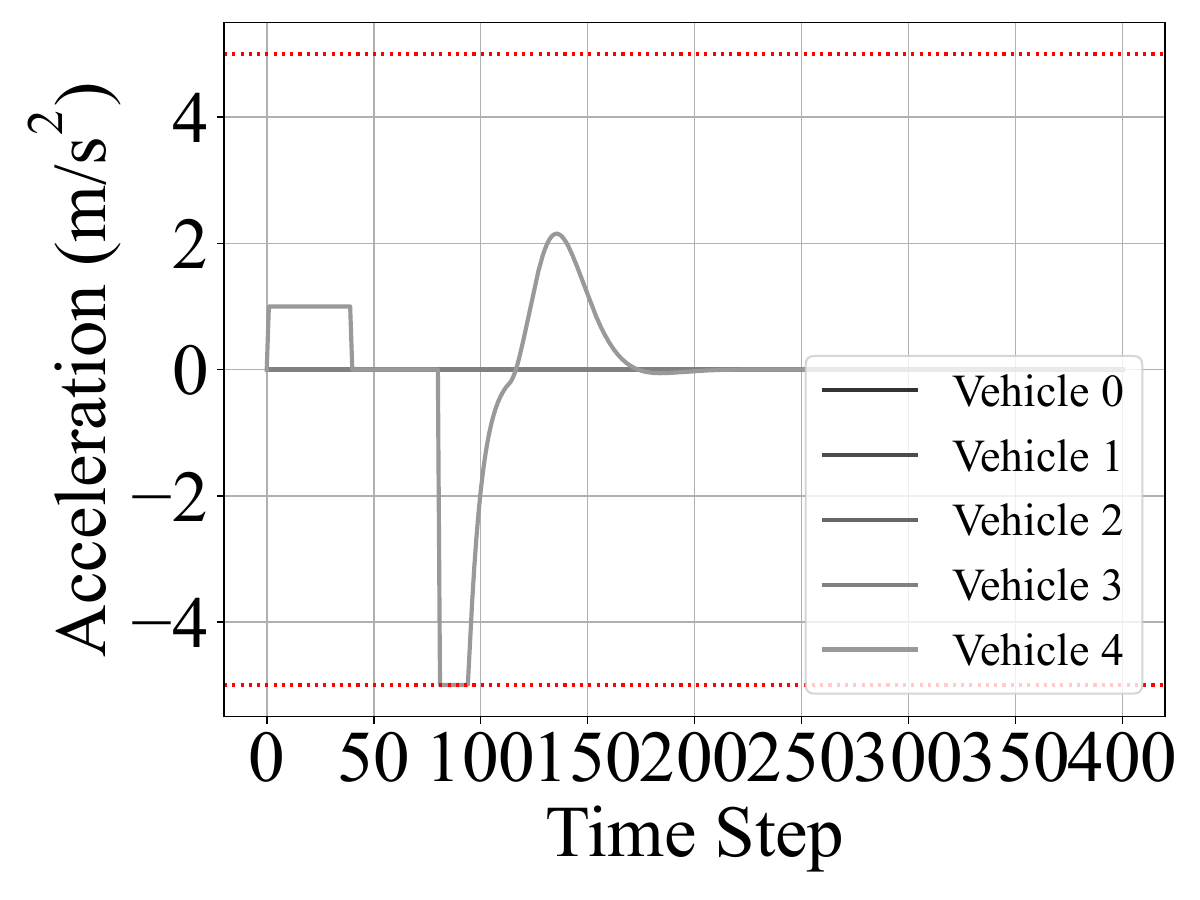}
        }
    \caption{Spacing, velocity and acceleration for vehicles in scenario 2 when vehicle 5 accelerates. (a), (b), (c) are the results for safe-RL with SI. (d), (e), (f) show the results for safe-RL without SI. (g), (h), (i) correspond to the results for PPO without safety guarantee. The results for the pure car-following scenario are presented in (j), (k), (l).}
    \label{fig: States for scenario 2-2}
\end{figure*}

We add disturbances to the $3$-rd or $4$-th HDV in the platoon, which is behind the CAV. At time $0$ s, the following HDV accelerates with $\rm 1 m/s^2$ for $4$s, maintains high velocity for $4s$, and then decelerates to an equilibrium state. The results are presented in Fig. \ref{fig: HDV safety 1} -- Fig. \ref{fig: States for scenario 2-2}. 

In Fig. \ref{fig: States for scenario 2-1}, both pure car following and PPO without safety guarantee fail to prevent a collision when vehicle 3 accelerates to approach the preceding vehicle. However, collisions are successfully avoided by safe-RL with SI and safe-RL without SI. In this case, the safety impact of disturbances resulting from inaccurate dynamics is minimal. Given that the safety constraints for the following HDVs are treated as soft constraints with relaxation coefficients in the optimization formulation \eqref{eq: control optimization}, it should be noted that the assurance for the value of the CBF candidate for the following HDVs within the safe set is not guaranteed but rather enhanced, as shown in Fig.~\ref{fig: HDV safety 1}.

In Fig. \ref{fig: States for scenario 2-2}, if vehicle $4$ accelerates, collisions occur when applying pure car following, PPO without safety guarantee, or safe-RL without SI. In contrast, safe-RL with SI can ensure safety, showcasing the importance of a more accurate estimation of the system dynamics to improve safety. Similarly, the value of the CBF candidate for the following vehicles is enhanced by safe-RL with SI, as shown in Fig.~\ref{fig: HDV safety 2}.

\section{Conclusion and Future Work}
\label{Conclusion}
In this paper, we propose a safe RL-based controller for the mixed-autonomy platoon by integrating the DRL method with a differentiable CBF layer to achieve system-level driving safety. A learning-based human driver behavior identification is utilized to derive the unknown dynamics of surrounding HDVs in mixed traffic scenarios. Traffic efficiency and string stability are taken into consideration by constructing comprehensive reward functions. The simulation results show that our proposed method effectively enhances traffic efficiency and stability while providing system-level safety guarantees in a mixed platoon. 
Furthermore, the safety layer contributes to expediting training by confining the range of exploration, which can be beneficial to the development and deployment of reinforcement learning algorithms. 

This research opens several promising directions for future work. First, we would like to apply a decision transformer to adapt to dynamic traffic scenarios where the number of surrounding vehicles of the CAV is time-varying. Second, it would be interesting to leverage multi-agent RL to extend the proposed framework to scenarios where multiple CAVs collaborate within a mixed-autonomy platoon to improve traffic flow, string stability, and system-level safety. Third, we would like to employ meta-RL to improve the generalizability of the proposed safe-RL algorithm to various unseen environments (e.g., road conditions, weather conditions, preferences over multiple control objectives, etc.). Fourth, the proposed methodology has the potential to be extended to consider other types of roads, such as (i) complex scenarios like merges, diverges, and weaving sections that require accounting for lane changes and (ii) urban roads that require coordination between platoon control and signal control. 

\bibliographystyle{IEEEtran}
\bibliography{ref.bib}

\begin{IEEEbiography}
[{\includegraphics[width=1in,height=1.25in,clip,keepaspectratio]{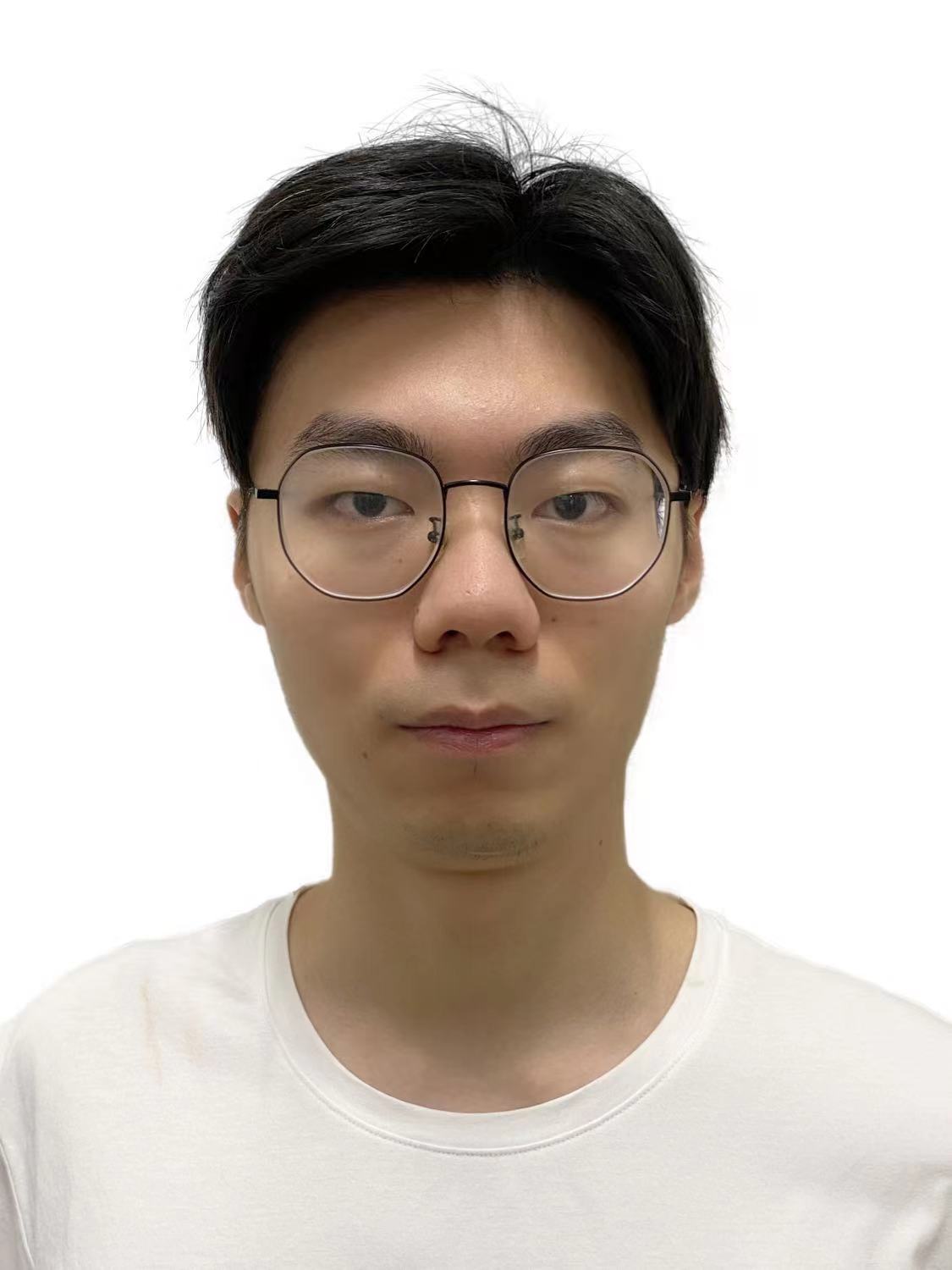}}]{Jingyuan Zhou} receives the B.Eng. degree in Electronic Information Science and Technology from Sun Yat-sen University, Guangzhou, China, in 2022. He is currently working towards a Ph.D. degree with the National University of Singapore. His research interests include safety-critical control and privacy computing of mixed-autonomy traffic.
\end{IEEEbiography}
\begin{IEEEbiography}
[{\includegraphics[width=1in,height=1.25in,clip,keepaspectratio]{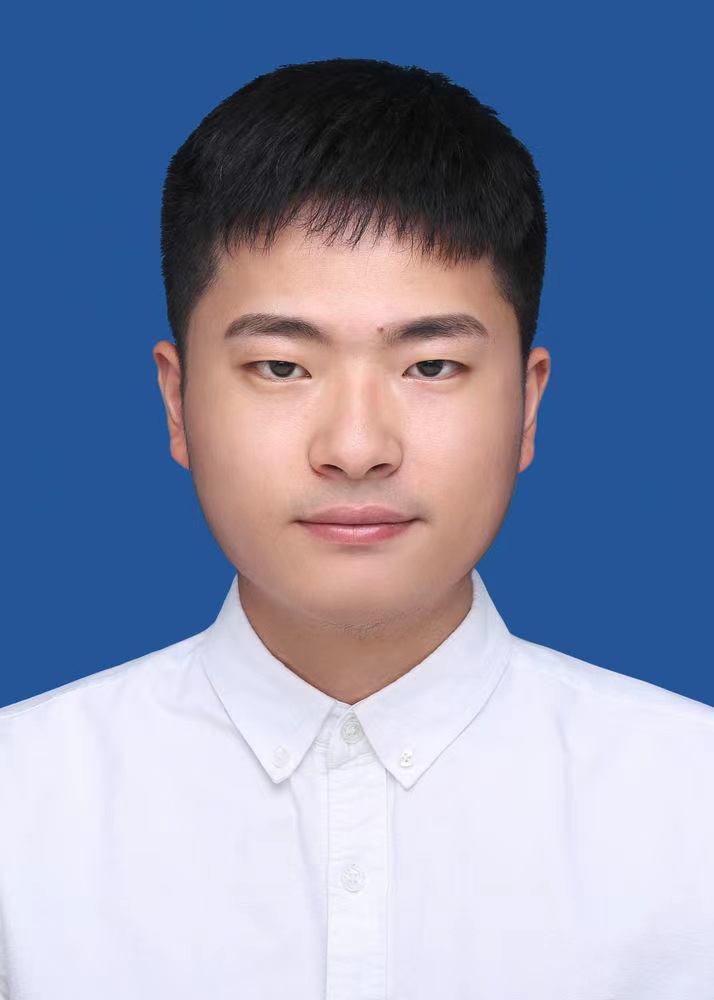}}]{Longhao Yan} receives the B.Eng. degree and M.Eng. degree in School of Electronics and Control Engineering from Chang’an University, Xi’an, China, in 2019 and 2022 respectively. He is currently working towards a Ph.D. degree with the National University of Singapore. His research interests include lateral control and trajectory prediction of intelligent transportation systems.
\end{IEEEbiography}
\begin{IEEEbiography}
[{\includegraphics[width=1in,height=1.25in,clip,keepaspectratio]{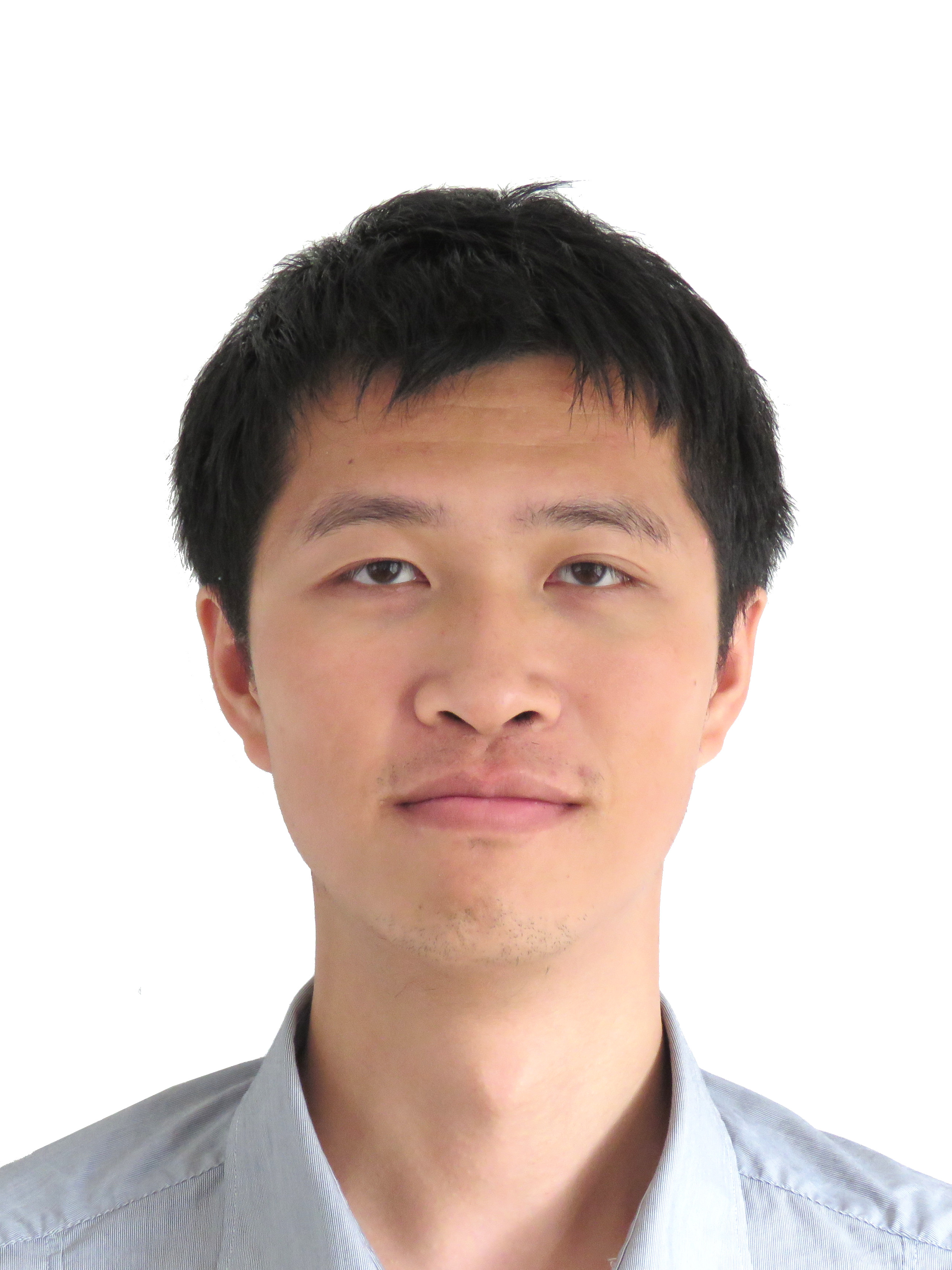}}]{Kaidi Yang}  is an Assistant Professor in the Department of Civil and Environmental Engineering at the National University of Singapore. Prior to this, he was a postdoctoral researcher with the Autonomous Systems Lab at Stanford University. He obtained a PhD degree from ETH Zurich and M.Sc. and B.Eng. degrees from Tsinghua University. His main research interest is the operation of future mobility systems enabled by connected and automated vehicles (CAVs) and shared mobility.
\end{IEEEbiography}
\vfill
\end{document}